\documentclass[aps,prr,twocolumn,superscriptaddress]{revtex4-2}

\usepackage{amssymb}
\usepackage{amsmath}
\usepackage{graphicx}
\usepackage[capitalise]{cleveref}
\usepackage{epsf}
\usepackage{tikz}
\usepackage{epsfig}
\usepackage{epstopdf}
\usepackage{color}
\usepackage{color}
\usepackage{float}
\usepackage{comment}

\def\g{{\bf{g}}}

\def\g0{{\gamma_0}}

\def\imp{{\mbox{\scriptsize imp}}}
\usepackage{cancel}
\usepackage[normalem]{ulem}
\newcommand{\blue}[1]{{\color{blue}{#1}}}

\def\figwidth{0.9\columnwidth}

\begin{document}
\title{Mobile Impurity in a Two-Leg Bosonic Ladder}

\author{Naushad Ahmad Kamar}
\affiliation{DQMP, University of Geneva, 24 Quai Ernest-Ansermet, 1211 Geneva, Switzerland}
\affiliation{Department of Physics and Astronomy, Michigan State University, East Lansing, Michigan 48824, USA}
\author{Adrian Kantian}
\affiliation{Department of Physics and Astronomy, Uppsala University, Box 516, S-751 20 Uppsala, Sweden}
\affiliation{Heriot-Watt University, Edinburgh, United Kingdom}
\author{Thierry Giamarchi}
\affiliation{DQMP, University of Geneva, 24 Quai Ernest-Ansermet, 1211 Geneva, Switzerland}
\begin{abstract}
We study the dynamics of a mobile impurity in a two-leg bosonic ladder. The impurity moves both along and across the legs and interacts with a bath of interacting bosonic particles present in the ladder.
We use both analytical (Tomonaga-Luttinger liquid - TLL) and numerical  (Density Matrix Renormalization Group - DMRG) methods to compute the Green's function of the impurity. We find that for a small impurity-bath interaction, the symmetric mode of the impurity effectively couples only to the gapless mode of the bath while the antisymmetric mode of the impurity couples to both gapped and gapless mode of the bath. We compute the time dependence of the Green's function of the impurity, for impurity created either in the antisymmetric or symmetric mode with a given momentum. The latter case leads to a decay as a power law below a critical momentum and exponential above, while the former case exhibits both power law and exponential decay depending on the transverse tunneling of the impurity. We compare the DMRG results with analytical results using the linked cluster expansion and find good agreement. In addition we use DMRG to extract the lifetime of the quasi-particle, when the Green's function decays exponentially. 
We also treat the case of an infinite bath-impurity coupling for which both the symmetric and the antisymmetric modes
are systematically affected. For this case the impurity Green's function in the symmetric mode decays as a power-law at zero momentum. The corresponding exponent increases with increasing transverse-tunneling of the impurity. 
We compare our results with  the other impurity problems for which the motion of either the impurity or the bath 
is limited to a single chain. Finally, we comment on the consequences of our findings for experiments with the ultracold gases.
\end{abstract}
\maketitle

\section{Introduction}
In a high dimensional bath, a mobile impurity behaves as a free particle,  with a renormalized mass and lifetime, this description of the impurity is known as quasi-particle (QP)~\cite{feynman55_polaron,feynman_statmech,caldeira_leggett,leggett_two_state}. The QP description has been successfully applied to many problems from condensed matter to ultracold gases~\cite{Franchini_Polaron_in_Materials,Jean_Dalibard_Fermionic_Atoms_Quasi_Particle,George_Bruun_Bose_Polaron}.  One classic example is the motion of an electron in the bath of phonons where the mass of the electron renormalizes and the electron behaves like a QP, known as polaron.

However it is known that several mechanisms can lead to a very different physics than simple quasiparticles. 
This is the case in the celebrated X-ray edge problem where the appearance of a static impurity induces an infinite number of excitations
in the bath leading to the famous Anderson orthogonality catastrophe \cite{Anderson_Orthogonality_Catastrophe,Giamarchi_bosonization}. Originally, the X-ray edge problem was observed for fermionic bath, it can also be extended to bosonic bath \cite{Giamarchi_bosonization}.
Similar physics occurs also in the Caldeira-Leggett problem where coupling to a bath can impede the tunneling of a macroscopic quantum variable \cite{caldeira_leggett,leggett_two_state}. Recently similar phenomena were shown to drastically 
affect the motion of impurities moving in a one dimensional bath of quantum interacting particles, leading 
to a motion quite different from a QP with a renormalized mass, namely to subdiffusion and the Green's function 
of the impurity exhibiting a power-law decay for a wide range of momenta~\cite{zvonarev_ferro_cold,kamenev_exponents_impurity,matveev_exponent_ferro_liquid,schecter_mobile_impurity,Lamacraft_mobile_impurity_in_one_dimension,zvonarev_bosehubbard,zvonarev_gaudinyang}. 
A part of this physics is due to the fact that in one dimension (1D) the recoil due to the motion of the 
impurity does not totally suppress Anderson orthogonality catastrophe contrarily to what happens in 
higher dimensions \cite{PhysRevB.3.1102,rosch1995heavy}. Thus one of the questions of interest is how a mobile impurity will behave in a bath which has both transverse and horizontal extensions. This is a first step towards studying the dimensional crossover in the impurity dynamics. To answer these questions, the dynamics of a mobile impurity in the ladder bath has been recently investigated 
for a system for which the impurity moves only along the legs of the ladder~\cite{Naushad_Ladder_Impurity} and in two decoupled chains where an impurity tunnels in both longitudinal and transverse directions \cite{Capone_Motion_of_an_impurity_in_decoupled_chain,Capone_Decoupled_chain_density_measurement}. For such systems, the impurity exhibits a similar class of dynamics as that of the 1D bath, but the power-law exponent becomes smaller in comparison to the one-dimensional bath. 

The study of a mobile impurity in a quantum bath is not limited to theoretical studies,  and experiments on the ultracold gases ~\cite{Michael_Kohl_Spin_impurity_in_bose_gas,Minardi_Dynamics_of_impurities_in_one_dimension,Fukuhara_spin_impurity_in_one_dimension,meinert_bloch_oscillations_TLL} provide in particular a platform to investigate such problem with a large flexibility and control on the impurity and the bath.  
  
In this work, we address the dynamics of a mobile impurity in a two-leg bosonic ladder with the impurity being able to tunnel between the two legs. Compared to the single chain case,  where the impurity 
was restricted to a 1D motion, we can expect that, in the present setup, the recoil effects could be more pronounced than in the strictly 1D case~\cite{Kopp:1990kn}. Another way to study such a problem is to consider the 
leg index as some ``spin'' index both for the bath and the impurity. In such a description the present 
problem would be a generalization of the Kondo problem (by opposition to the X-ray edge one with a 
featureless impurity) but with the possibility of motion of the impurity. This poses the question 
of the subtle coupling of the internal and center of mass degrees of freedom.

We study this problem using the numerical method time-dependent density matrix renormalization group (t-DMRG)~\cite{white_dmrg_letter,Schollwock_DMRG} and analytical methods such as Tomonaga-Luttinger liquid (TLL)~\cite{haldane_Luttinger2,Giamarchi_bosonization} and linked cluster expansion (LCE)~\cite{Mahan_Many_Particle_Physics}. The t-DMRG allows us to access the impurity dynamics from weak to strong interactions with the bath, while LCE describes the impurity dynamics in the weak coupling limit. We compare our results with previous studies on the impurity dynamics in the one-dimensional bath~\cite{Adrian_Kantian_Mobile_Impuirty} and the ladder bath~\cite{Naushad_Ladder_Impurity}.    

The plan of the paper is as follows: In Sec.~\ref{sec:definitions}, we describe the model on the lattice and in the continuum limit, its bosonization representation, and various observables. In Sec.~\ref{sec:analytic}, we describe the analytical expression of the observables by using bosonization and the LCE. In Sec.~\ref{sec:numerics}, we present the numerical t-DMRG~\cite{white_dmrg_letter,Schollwock_DMRG}, analysis of this problem, and the results for the Green's function of the impurity. Sec.~\ref{sec:discussion}, discusses these results both in connection with the one-dimensional motion of an impurity in a ladder's results and in view of the possible extensions. Finally, Sec.~\ref{sec:conclusion} concludes the paper and presents some perspectives in connection with experiments.  
The analytical expression of the Green's function is given in the appendix \ref{ap:LCE}.
\section{Mobile Impurity in a Two Leg Bosonic Ladder} \label{sec:definitions}
\subsection{Model} \label{sec:model}
\begin{figure}
\begin{center}
 \includegraphics[width=\figwidth]{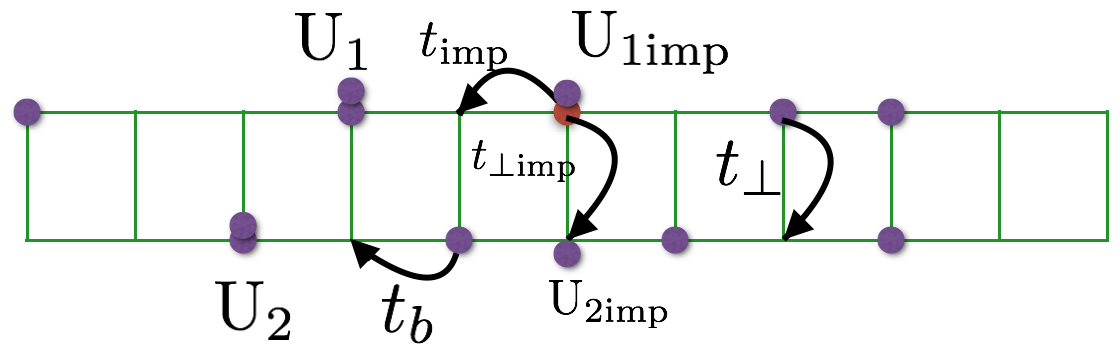}
 \end{center}
\caption{\label{fig:2fig1}
(color online) Impurity in a two-leg  bosonic ladder. The blue solid circles represent the bath particles and the red circle represents the impurity.
 The bath particles move along the legs (resp. between the legs) with hopping $t_b$ (resp. $t_\perp$)(see text).
 The impurity moves in both longitudinal and transverse directions with  amplitudes $t_{\text{imp}}$ and $t_{\perp \text{imp}}$ (see text), respectively. The impurity and the bath particles interact by  the contact interactions $U_{1\text{imp}}$ and $U_{2\text{imp}}$ in leg 1 and leg 2, respectively.}
\end{figure}

We consider a mobile impurity moving in a two-leg Bosonic ladder in both horizontal and transverse directions.
The model we consider is depicted in Fig.~\ref{fig:2fig1}.

The full Hamiltonian is given by
\begin{eqnarray} \label{eq:hamdiscrete}
 H  = H_K + H_{\text{lad}} + U_{1\text{imp}} \sum_{j=1}^L \rho_{1,j} \rho_{\text{imp},1,j}\\\nonumber+ U_{2\text{imp}} \sum_{j=1}^L \rho_{2,j} \rho_{\text{imp},2,j},
\end{eqnarray}
where $U_{1\text{imp}}$, $U_{2\text{imp}}$, and $L$ are the interaction strength between the particles in the leg 1, in the leg 2  with  the impurity, and the ladder size along longitudinal direction, respectively. Also, $\rho_{1,j}$, $\rho_{2,j}$, $\rho_{\text{imp},1,j}$, and $\rho_{\text{imp},2,j}$ denote the density operator of the bath in leg 1, leg 2, the density operator of the impurity in leg 1, and in leg 2, respectively. We consider $U_{1\text{imp}}=U_{2\text{imp}}=U$.

The impurity kinetic energy  is given by the tight-binding Hamiltonian
\begin{multline}
 H_K = - t_{\text{imp}} \sum_{j=1}^{L-1} (d^\dagger_{1,j+1} d_{1,j} +d^\dagger_{2,j+1} d_{2,j}+ \text{h.c.}) \\ 
 -t_{\perp\text{imp}} \sum_{j=1}^{L} (d^\dagger_{1,j} d_{2,j} + \text{h.c.}).
\end{multline}
We diagonalize $H_K$ by using  symmetric and antisymmetric combinations of $d_{1,j}$ and $d_{2,j}$, and $H_K$ can be re-expressed as 
\begin{equation}
H_{K} = \sum_{q}\epsilon_s(q) d^\dagger_{s,q}d_{s,q}+\epsilon_a(q) d^\dagger_{a,q}d_{a,q} 
\end{equation}
where $d_{\gamma, j}$ ($d^\dagger_{\gamma, j}$) are the destruction (creation) operators of the impurity on site $j$ in leg $\gamma=1,2$, $d_{s,q}=\frac{d_{1,q}+d_{2,q}}{\sqrt 2}, d_{a,q}=\frac{d_{1,q}-d_{2,q}}{\sqrt 2}$, $\epsilon_a(q)=-2t_{\text{imp}}\cos(q)+t_{\perp\text{imp}}$, and $\epsilon_s(q)=-2t_{\text{imp}}\cos(q)-t_{\perp\text{imp}}$.
The density of the impurity $\rho_{\text{imp},\gamma,j}$ and Fourier transformation of $d_{\gamma, j}$ are defined as
\begin{equation}
\begin{split}
 \rho_{\text{imp},\gamma,j} &= d^\dagger_{\gamma,j} d_{\gamma,j},\\
 \hat{d}_{\gamma,q} &= \sum_j e^{i q r_j} d_{\gamma,j},
 \end{split}
\end{equation}
where $r_j=a j$, and $a=1$ is lattice constant.
The ladder Hamiltonian $H_{\text{lad}}$ is given by
\begin{equation} \label{eq:laddiscrete}
 H_{\text{lad}} = H^0_1 + H^0_2 - t_\perp \sum_{j=1}^{L} (b^\dagger_{1,j} b_{2,j} + \text{h.c.} )
\end{equation}
where $b_{\gamma,j}$ ($b^\dagger_{\gamma,j}$) are the destruction (creation) operators for a boson in the bath on chain $\gamma$ and site $j$.
The operator $b$ obeys the usual bosonic commutation relation rules.
The single chain Hamiltonian is the
Bose-Hubbard one
\begin{equation}
 H^0_i = -t_b \sum_{j=1}^{L-1} (b^\dagger_{i,j+1} b_{i,j} + \text{h.c.}) + \frac{U_i}2 \sum_{j=1}^{L} \rho_{i,j}(\rho_{i,j} - 1) - \mu_i \sum_j \rho_{i,j},
\end{equation}
where $\rho_{\gamma,j}=b^\dagger_{\gamma,j}b_{\gamma,j}$ is density operator of the ladder in leg $\gamma$ and site $j$. The eq.~(\ref{eq:hamdiscrete}) is convenient for the numerical study. In order to make connection with the field theory analysis we can also consider the same problem
in a continuum. In that case the Hamiltonian becomes
\begin{multline}\label{eq:hamtot}
 H = \frac{P^2}{2M} - t_{\perp\text{imp}}(|1\rangle\langle 2|+|2\rangle\langle 1|) \\+
  H_{\text{lad}} + U (\rho_1(X)|1\rangle\langle 1|+\rho_2(X)|2\rangle\langle 2|),
\end{multline}
where $X$, $P$ are the position, momentum operators of the impurity , and $\rho_\gamma(X)$ is density operator of the bath at position $X$ in leg $\gamma$ of the ladder. The basis set $|1\rangle$ and $|2\rangle$ represent the chain index of the ladder, and they form a complete basis ($|1\rangle\langle 1|+|2\rangle\langle 2|=I$, where $I$ is a $2\times 2$ identity matrix). The last two terms in \cref{eq:hamtot} arise from the fact that the impurity density operators in leg 1 and leg 2 are $\rho_{\imp,1}(x)=\delta(x-X)|1\rangle\langle 1|$ and $\rho_{\imp,2}(x)=\delta(x-X)|2\rangle\langle 2|$, respectively. 

The ladder Hamiltonian (\ref{eq:laddiscrete})  in the continuum becomes 
\begin{equation} \label{eq:hamlad}
 H_{\text{lad}} = H^0_1 + H^0_2 - t_\perp \int dx (\psi^\dagger_1(x)\psi_2(x) + \text{h.c.}),
\end{equation}
and the single chain Hamiltonian is
\begin{equation} \label{eq:hamchain}
 H^0_i = \frac1{2m} \int dx |\nabla\psi_i(x)|^2 + \frac{U_i}2 \int dx \rho_i(x)^2 - \mu_i \int dx \rho_i(x),
\end{equation}
where $m$ is the mass of the bosons, $\mu_i$ is the chemical potential and $U_i$ is the intrachain interaction.
\subsection{Observables}
To characterize the dynamics of the impurity in the ladder we mostly focus on the Green's function of the impurity. We study it at zero temperature both analytically and numerically via DMRG.
Compared to the case \cite{Naushad_Ladder_Impurity}  where the impurity was confined to a single chain, it is now necessary to introduce two independent Green's functions for the impurity. 
The Green's functions are in the chain basis
\begin{equation}\label{eq:greenimp}
G_{\alpha\beta}(p,t)= \langle \hat{d}_{\alpha,p}(t) \hat{d}_{\beta,p}^\dagger(t=0) \rangle,
\end{equation}
where $\alpha$ and $\beta$ can take the values $1,2$ corresponding to the chain index and $\langle \cdots \rangle$ denotes the average in the ground state of the bath, and with zero impurities present. $O(t)$ denotes the usual Heisenberg time evolution of the operator
\begin{equation}
 O(t) = e^{i H t} O e^{-i H t}.
\end{equation}
By symmetry 
we can restrict ourselves to $G_{11}(p,t)$ and  $G_{12}(p,t)$. The two other Green's functions are simply related to (\ref{eq:greenimp}) by $G_{22}(p,t)=G_{11}(p,t)$ and $G_{12}(p,t)=G_{21}(p,t)$,

Instead of using the chain basis it can 
be more convenient to use the symmetric and antisymmetric  operators of the impurity,leading to the two Green's 
functions
\begin{equation} \label{eq:greenimpbonding}
\begin{split}
 G_{s}(p,t) &= \langle \hat{d}_{s,p}(t) \hat{d}_{s,p}^\dagger(t=0) \rangle,\\
 G_{a}(p,t) &= \langle \hat{d}_{a,p}(t) \hat{d}_{a,p}^\dagger(t=0) \rangle.
 \end{split}
\end{equation}
 One has $G_{s}(p,t)=G_{11}(p,t)+G_{12}(p,t)$, $G_{a}(p,t)=G_{11}(p,t)-G_{12}(p,t)$. 
\subsection{Bosonization representation}
To deal with the Hamiltonian defined in the previous section, we use the fields $\theta_\alpha(x)$ and $\phi_\alpha(x)$ ~\cite{Giamarchi_bosonization} for chain $\alpha=1, 2$ which are related to the field operators of the system via
\begin{equation}\label{eq:2eq1c}
 \rho_\alpha(x) =  \rho_{0,\alpha}-\frac{\nabla \phi_{\alpha}(x)}{\pi}+\rho_{0,\alpha} \sum_{p\neq 0} e^{2 i p(\pi\rho_{0, \alpha} x-\phi_{\alpha}(x))},
\end{equation}
where $\rho_{0,\alpha}$ is the average density on the chain $\alpha=1,2$. For equivalent upper and lower legs of the ladder, we can take $\rho_{0,1}=\rho_{0,2}=\rho_0$ . The creation operator of a particle in the bath in terms of $\theta$ and $\phi$ is given to lowest order by
\begin{equation}\label{eq:2eq5}
 \psi_{\alpha}^{\dagger}(x) = \rho_{0,\alpha}^{1/2} e^{-i\theta_\alpha(x)}.
\end{equation}
The conjugate field operators $\phi_{1,2}$ and $\theta_{1,2}$ obey
\begin{equation}\label{eq:2eq3}
\Big[\phi(x_1),\frac{\nabla \theta(x_2)}{\pi}\Big] = i \delta(x_1-x_2).
\end{equation}

Using the bosonization framework, the Hamiltonian of the ladder is given by 
\begin{equation} \label{eq:ladcont}
 H_{\text{lad}} = H_s + H_a,
\end{equation}
with
\begin{equation}\label{eq:2eq7}
\begin{split}
H_s =& \frac{1}{2\pi}\int dx[u_s K_s(\partial_x \theta_s)^2+\frac{u_s}{K_s} (\partial_x \phi_s)^2 ], \\
H_a =& \frac{1}{2\pi}\int dx[u_a K_a(\partial_x \theta_a)^2+\frac{u_a}{K_a} (\partial_x \phi_a)^2 ]\\
     & -2\rho_0 t_\perp\int dx \cos(\sqrt2\theta_a(x)),
\end{split}
\end{equation}
and
\begin{equation}
\begin{split}
 \theta_{s,a} =& \frac{\theta_1\pm\theta_2}{\sqrt{2}},\\
 \phi_{s,a} =& \frac{\phi_1\pm\phi_2}{\sqrt{2}}.
\end{split}
\end{equation}
$K_{s,a}$ and $u_{s,a}$ are known as Luttinger parameters and control the properties of the bosonic ladder.
The cosine term \cite{Giamarchi_bosonization,Naushad_Ladder_Impurity} opens a gap $\Delta_a$ in the antisymmetric sector when $K_a > 1/4$. This massive phase for the antisymmetric mode signals the existence of phase 
coherence across the two legs of the ladder, with exponentially decreasing correlations for the antisymmetric density-density correlations.  
The symmetric sector is described by the usual TLL Hamiltonian, and has power-law correlations. 
A numerical calculation of the TLL parameters for the massless phase can be found in Ref. \cite{crepin_bosonic_ladder_phase_diagram}.
\section{Analytical solution for weak coupling ($U<<\Delta_a$)} \label{sec:analytic}
Let us now investigate the full Hamiltonian (\ref{eq:hamtot}) (or (\ref{eq:hamdiscrete})) to compute the Green's function of the impurity (\ref{eq:greenimp}).

Using (\ref{eq:2eq1c}) the interaction term $H_{\text{coup}}$ with the impurity  in terms of the $d_s$ and $d_a$ becomes
\begin{multline}
 H_{\text{coup}} = \frac{U}{2} \int dx (\rho_1(x)+\rho_2(x)) (d_s(x)^\dagger d_s(x)+ d_a(x)^\dagger d_a(x))\\  
 +\frac{U}{2} \int dx (\rho_1(x)-\rho_2(x)) (d_s(x)^\dagger d_a(x)+h.c ),
\end{multline}
which leads to the expression, in terms of the symmetric and antisymmetric modes of the bath,

\begin{multline} \label{eq:hambossym_0}
 H_{\text{coup}} = \frac{-U}{\sqrt 2 \pi} \int dx \nabla\phi_s(x) (d_s(x)^\dagger d_s(x)+ d_a(x)^\dagger d_a(x)) \\  -\frac{U}{\sqrt 2 \pi} \int dx \nabla\phi_a(x) (d_s(x)^\dagger d_a(x)+h.c ) \\
 +\frac{U\rho_0}{2}\int dx \cos(2\pi\rho_0 x-\sqrt 2 \phi_s)\cos(\sqrt 2 \phi_a)\\ \times (d_s(x)^\dagger d_s(x)+ d_a(x)^\dagger d_a(x))\\-\frac{U\rho_0}{2}\int dx\sin(2\pi\rho_0 x-\sqrt 2 \phi_s)\sin(\sqrt 2 \phi_a)\\ \times (d_s(x)^\dagger d_a(x)+h.c ) .
\end{multline}
Since the field $\theta_a$ is ordered for $K_a>1/4$, the correlation functions corresponding to its dual field $\phi_a$ will be exponentially suppressed. Thus for $U\ll \Delta_a$, we can drop the cosine and sine terms in (\ref{eq:hambossym_0}) and finally in weak coupling regime, (\ref{eq:hambossym_0}) can be expressed as 

\begin{multline} \label{eq:hambossym}
 H_{\text{coup}} = \frac{-U}{\sqrt 2 \pi} \int dx \nabla\phi_s(x) (d_s(x)^\dagger d_s(x)+ d_a(x)^\dagger d_a(x)) \\  -\frac{U}{\sqrt 2 \pi} \int dx \nabla\phi_a(x) (d_s(x)^\dagger d_a(x)+h.c ) .
\end{multline}
Note that we have kept in \cref{eq:hambossym} only the most relevant term, which for bosons  is the 
forward scattering on the symmetric and antisymmetric modes of the bath. In \cref{eq:hambossym}, the impurity-bath coupling $U$ is distributed on both symmetric and antisymmetric modes of the bath with an effective interaction $U/\sqrt{2}$.

To compute the Green's functions, we use the same approach as in \cite{Naushad_Ladder_Impurity} namely,
the linked cluster expansion (LCE) \cite{Kopp:1990kn,zvonarev_ferro_cold, Adrian_Kantian_Mobile_Impuirty}.
The calculation is detailed in appendix \ref{ap:LCE} and gives the asymptotic behavior of the impurity Green's function (\ref{eq:greenimp}) for $2t_{\perp\text{imp}}>\frac{\Delta_a\sqrt{2 u_a \pi}}{\sqrt{K_a}}$ as
\begin{equation} \label{eq:2eq30}
\begin{split}
|G_{s}(0,t)| &\simeq \Big(\frac{1}{t}\Big)^\alpha ,\\
|G_{a}(0,t)| &\simeq e^{-A_2 U^2 \pi t} \Big(\frac{1}{t}\Big)^\alpha,
\end{split}
\end{equation}
and for $2t_{\perp\text{imp}}<\frac{\Delta_a\sqrt{2 u_a \pi}}{\sqrt{K_a}}$
\begin{equation} \label{eq:2eq30_new}
\begin{split}
|G_{s}(0,t)| &\simeq \Big(\frac{1}{t}\Big)^\alpha ,\\
|G_{a}(0,t)| &\simeq \Big(\frac{1}{t}\Big)^\alpha,
\end{split}
\end{equation}

where $\alpha=\frac{K_s U^2}{4 \pi^2 u_s^2}$, and $K_s=.835$, $u_s=1.86$, for $t_b=t_\perp=1, U_1=U_2=\infty, \rho_0=1/3$ \cite{crepin_bosonic_ladder_phase_diagram} and 
\begin{align}
A_2&\simeq\frac{ K_a}{4 u_a \pi^2}\frac{(u_a^2q_-^2+\tilde{\Delta}^2)}{ q_-(2t_{\text{imp}}\sqrt{u_a^2q_-^2+\tilde{\Delta}^2}+u_a^2)},
\end{align}
where $q_-$ and $\tilde{\Delta}$ are expressed in appendix \ref{ap:LCE}.
In our LCE calculation we also find that for $2t_{\perp\text{imp}}<\frac{\Delta_a\sqrt{2 u_a \pi}}{\sqrt{K_a}}$, Green's function in both symmetric and antisymmetric sectors decay as a power-law (see \cref{eq:2eq30}) at $p=0$. This is an emergent effect of the ladder bath and transverse tunneling $t_{\perp\text{imp}}$ on the dynamics of the impurity. For decoupled chains ($\Delta_a=0$)~\cite{Capone_Motion_of_an_impurity_in_decoupled_chain}, the impurity Green's function decays as power-law and exponentially in the symmetric and antisymmetric sector at any finite transverse tunneling of the impurity, respectively, this behaviour is also reproduced in \cref{eq:2eq30}.

For a weak repulsion between the impurity and the bath and  $2t_{\perp\text{imp}}>\frac{\Delta_a\sqrt{2 u_a \pi}}{\sqrt{K_a}}$ , we thus find that the Green's function in the symmetric mode  decays as a power-law with time as was the case with an impurity confined to a single chain \cite{Naushad_Ladder_Impurity}. In the antisymmetric mode on the other hand it decays exponentially.
\section{Numerical solution} \label{sec:numerics}
Analyzing the regime $U \gg \Delta_a$  is much more involved since now excitations across the antisymmetric gap can be created. We thus turn to a numerical analysis of this problem. 
\subsection{Method}
We use time-dependent DMRG (t-DMRG) \cite{Schollwock_density_matrix_renormalization_group} to compute the Green's function of the impurity, and we follow the method described in Ref.~\cite{Adrian_Kantian_Mobile_Impuirty, Naushad_Ladder_Impurity,kamar2019quantum_thesis}. For completeness let us recall the method, which is described below.

We map the ladder-impurity problem to a one-dimensional problem by a supercell approach. We denote bath particles in leg 1 and leg 2 by $B$ and $C$, the impurity in leg 1 and leg 2 by $A$ and $D$ and the total number of bath particles and number of impurity are conserved separately.  The local dimension of Hilbert  space for A, B, C, D is two for hardcore bosons hence the dimension of local  Hilbert space of supercell (A, B, C, D) is $2\times2\times2\times2=16$. We compute the Green's function of the impurity in the ground state of the ladder . The ground state $(|GS\rangle)$ is computed using DMRG. The Green's functions $G_{11}(x,t)(G_{12}(x,t))$ of the impurity in Heisenberg picture is given by 
\begin{align}
G_{11}(x,t)=e^{i E_{GS} t}\langle GS| d_{1, \frac{ L+1}{2}-x}e^{-i H t} d^\dagger_{1, \frac{L+1}{2}}|GS\rangle, \nonumber\\
G_{12}(x,t)=e^{i E_{GS} t}\langle GS| d_{2, \frac{ L+1}{2}-x}e^{-i H t} d^\dagger_{1, \frac{L+1}{2}}|GS\rangle,
\end{align}
where $E_{GS}$ is ground state energy of the bath. We compute $e^{-i H t} d^\dagger_{1, \frac{L+1}{2}}|GS\rangle$  using t-DMRG and $\langle GS| d_{2, \frac{ L+1}{2}-x}$ $(\langle GS| d_{1, \frac{ L+1}{2}-x})$ are computed using DMRG. 
By using $G_{11}(x,t)$ and $G_{12}(x,t)$, we compute  $G_s(x,t)=G_{11}(x,t)+G_{12}(x,t)$ and $G_a(x,t)=G_{11}(x,t)-G_{12}(x,t)$. 
For the numerical calculation we have used a bath of hardcore bosons at a density of $\rho_0=1/3$. This choice avoids the Mott-insulating phase that the ladder's symmetric mode might enters at commensurate density \cite{crepin_bosonic_ladder_phase_diagram}. We have used a bond dimension $\chi=400$ to compute the Green's function in a reasonable time. We chose Hamiltonian parameters  $t_\perp=t_b=t_{\text{imp}}=1, U_1=U_2=\infty$, and various values of $t_{\perp\text{imp}}$ and $U$. We fix the size of the system to $L=101$ sites per leg.

\begin{figure}
\begin{center}
  \includegraphics[ scale=0.35]{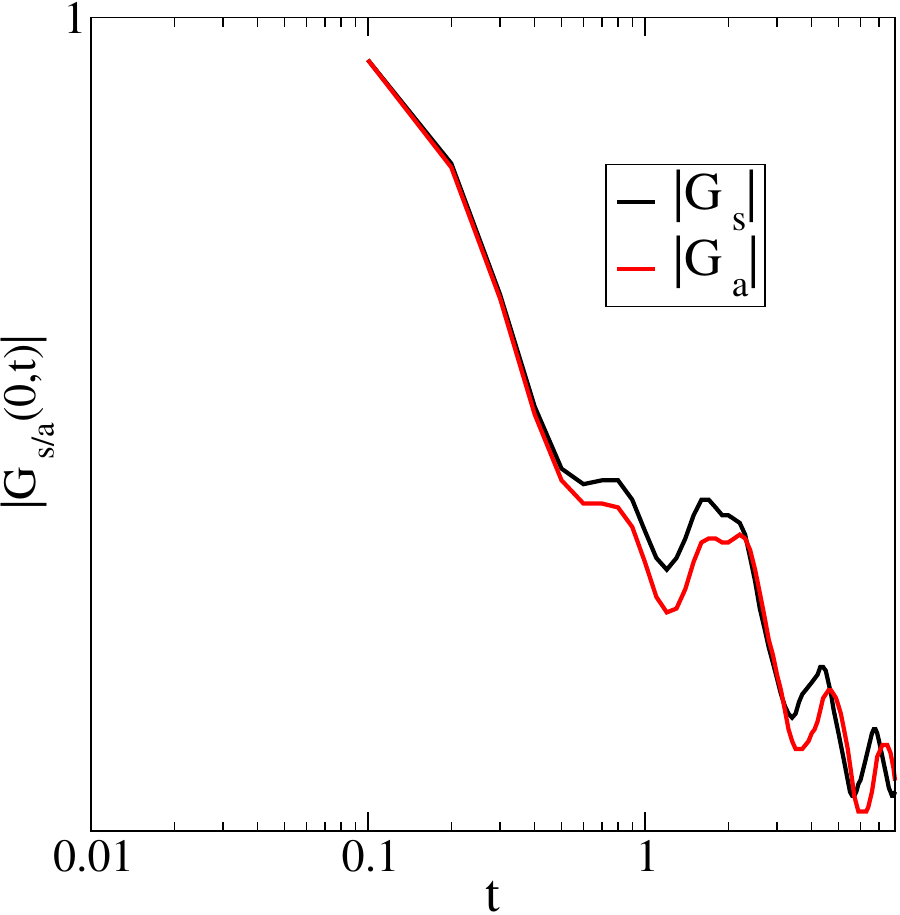}
  \includegraphics[ scale=0.35]{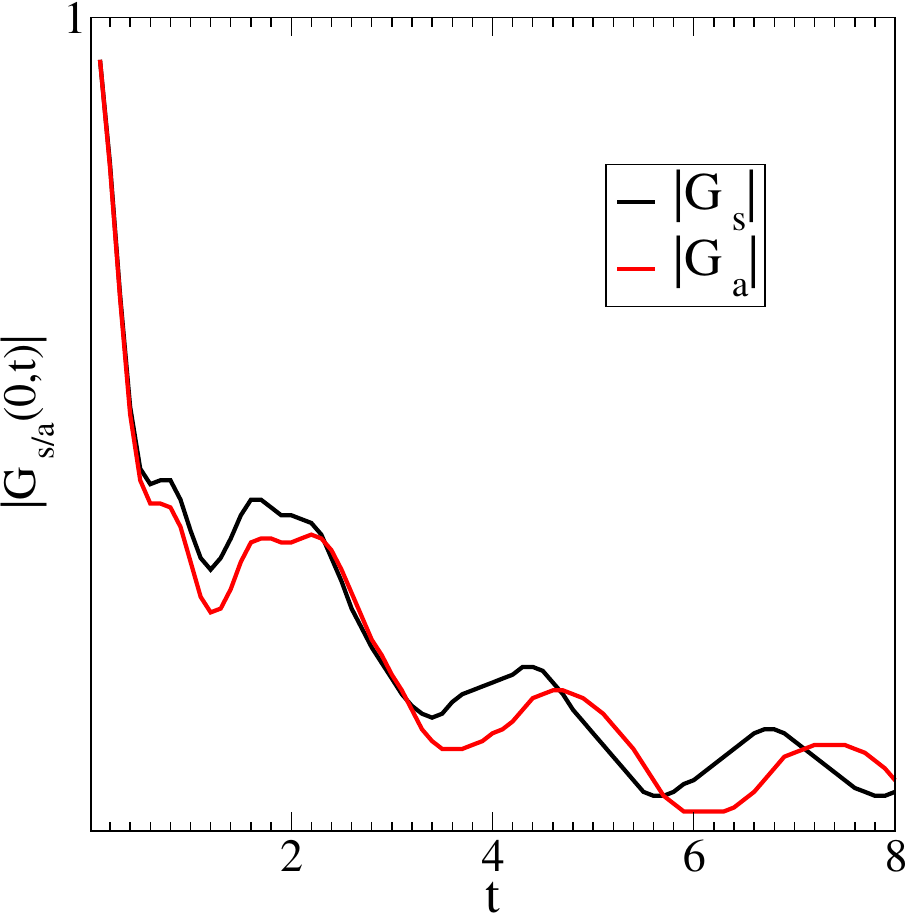}
\end{center}
\caption{\label{fig:sym_asym_green_function_t_0_5_U_1_0} (color online) Modulus of the impurity Green’s function at $p=0$ in the symmetric and antisymmetric sectors, shown on a log-log (upper panel) and semi-log (lower panel) scale as a function of time for a hardcore bosonic bath. 
Simulation parameters are $t_{\text{imp}} = t_{\text{b}} = 1$, $t_{\perp} = 1.0$, $U = 1.0$, $t_{\perp\text{imp}} = 0.5$, and momentum $p = 0$. 
At small $t_{\perp\text{imp}} = 0.5$, the Green’s function decays as a power law in both sectors.}
\end{figure}

\subsection{Zero momentum regime}\label{sec:zeromoment}

\begin{figure}
\begin{center}
  \includegraphics[ scale=0.4]{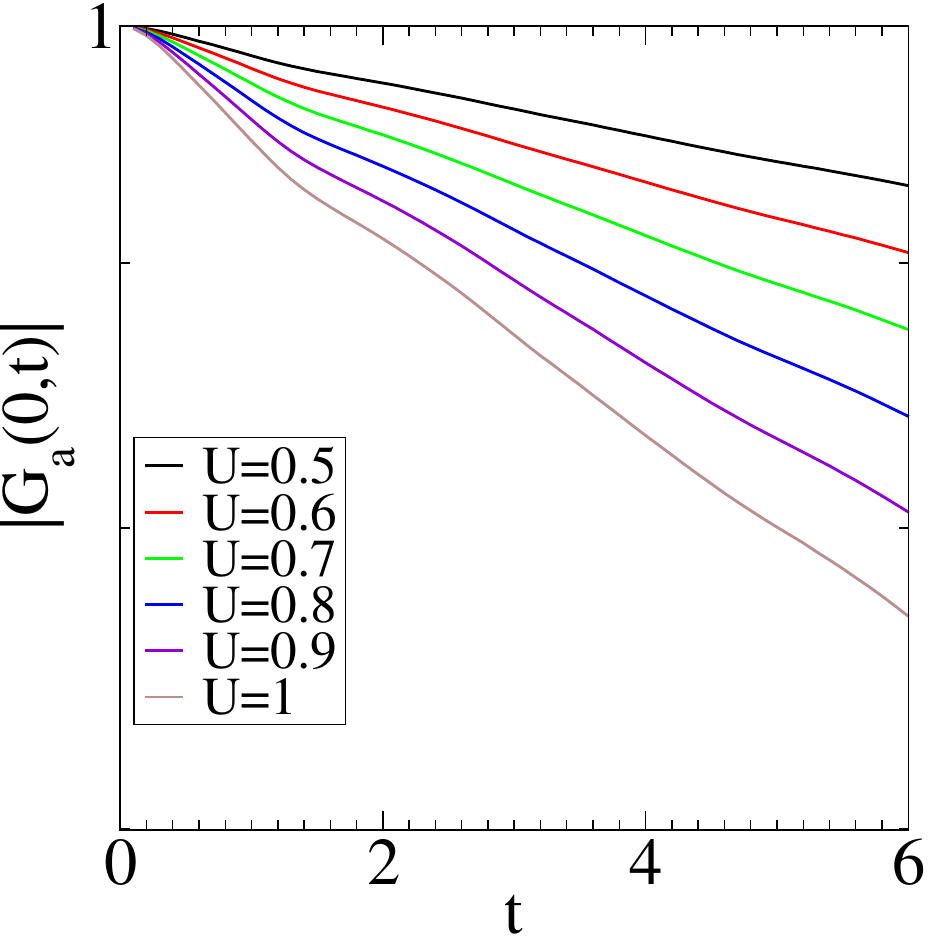}
   \includegraphics[ scale=0.4]{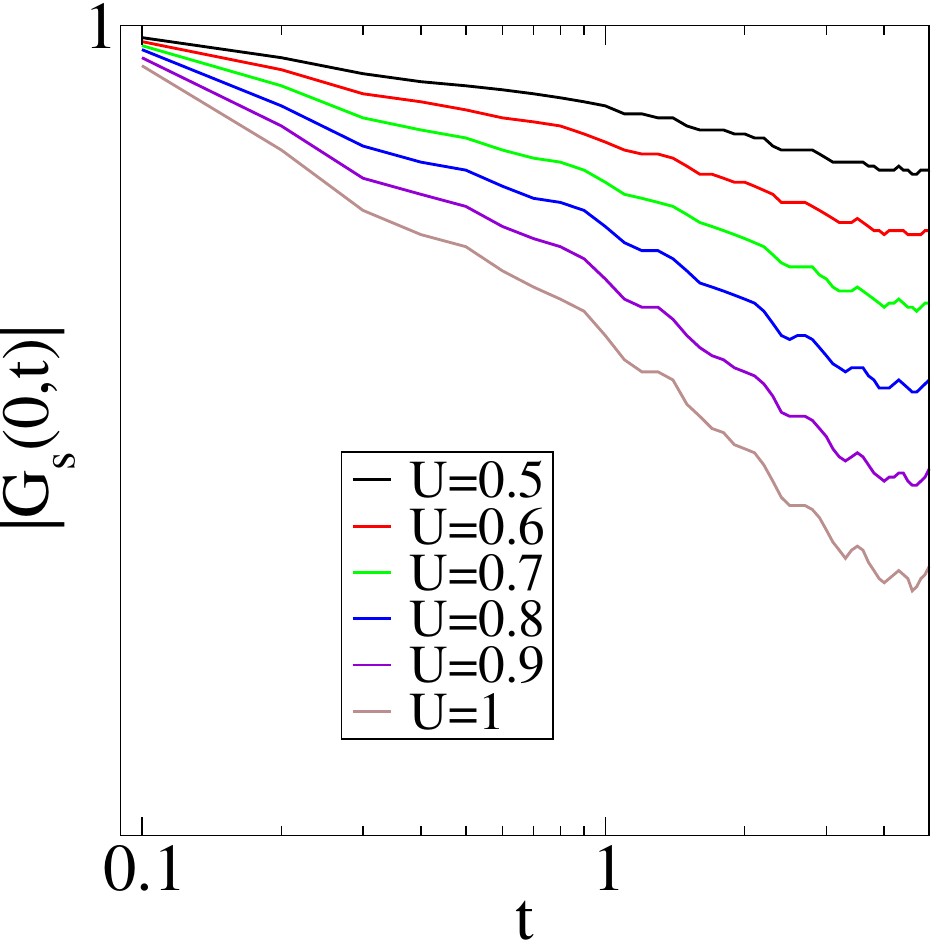}
\end{center}
\caption{\label{fig:sym_asym_green_function} (color online)   The modulus of the Green's function of the impurity in antisymmetric (upper panel) and symmetric sectors (lower panel)  as a function of time for hardcore bosonic bath at $t_\text{imp}= t_\text{b}=1$,  $\text{t}_\perp=1$, $\text{U}$ runs from $0.5$ to $1$, $t_{\perp\text{imp}}=3$ and $\text{p}=0$. The upper panel is depicted on semi-log scale shows a linear behavior while lower panel shows a linear behavior on log-log scale.  }
\end{figure}

We show  the Green's function of the impurity in the antisymmetric and symmetric modes $|G_a(p,t)|, |G_s(p,t)|$ at momentum $p=0$ in Fig.~\ref{fig:sym_asym_green_function_t_0_5_U_1_0} and Fig.~\ref{fig:sym_asym_green_function}  on semi-log and log-log scales, we find that $|G_s(0,t)|$ decays as a power-law which is similar to one observed in one-dimensional motion of the impurity in a two-leg bosonic ladder \cite{ Naushad_Ladder_Impurity}. However, the Green's function of the impurity  in the antisymmetric mode shows a power-law decay at small $t_{\perp \text{imp}}=0.5$ as shown in Fig.~\ref{fig:sym_asym_green_function_t_0_5_U_1_0}, which is in agreement with our LCE findings while for large $t_{\perp \text{imp}}=3.0$, it shows an exponential decay (shown in Fig.~\ref{fig:sym_asym_green_function}). 

\begin{figure}
\begin{center}
  \includegraphics[ scale=0.4]{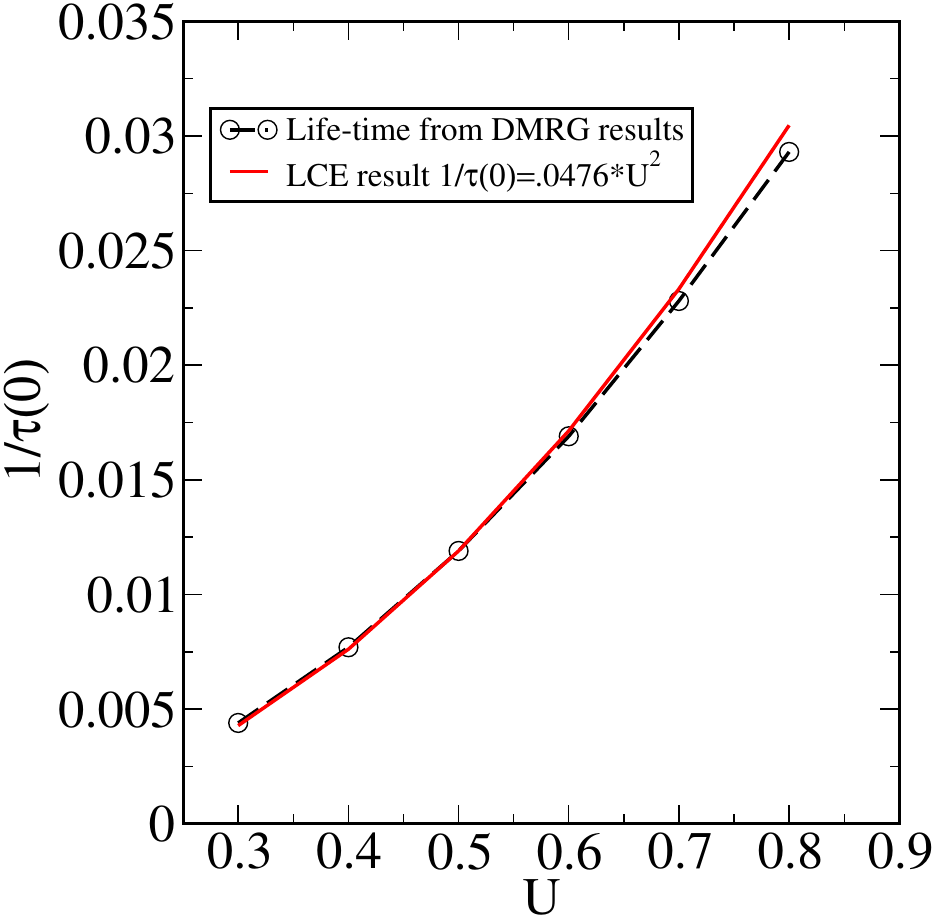}
   \includegraphics[ scale=0.4]{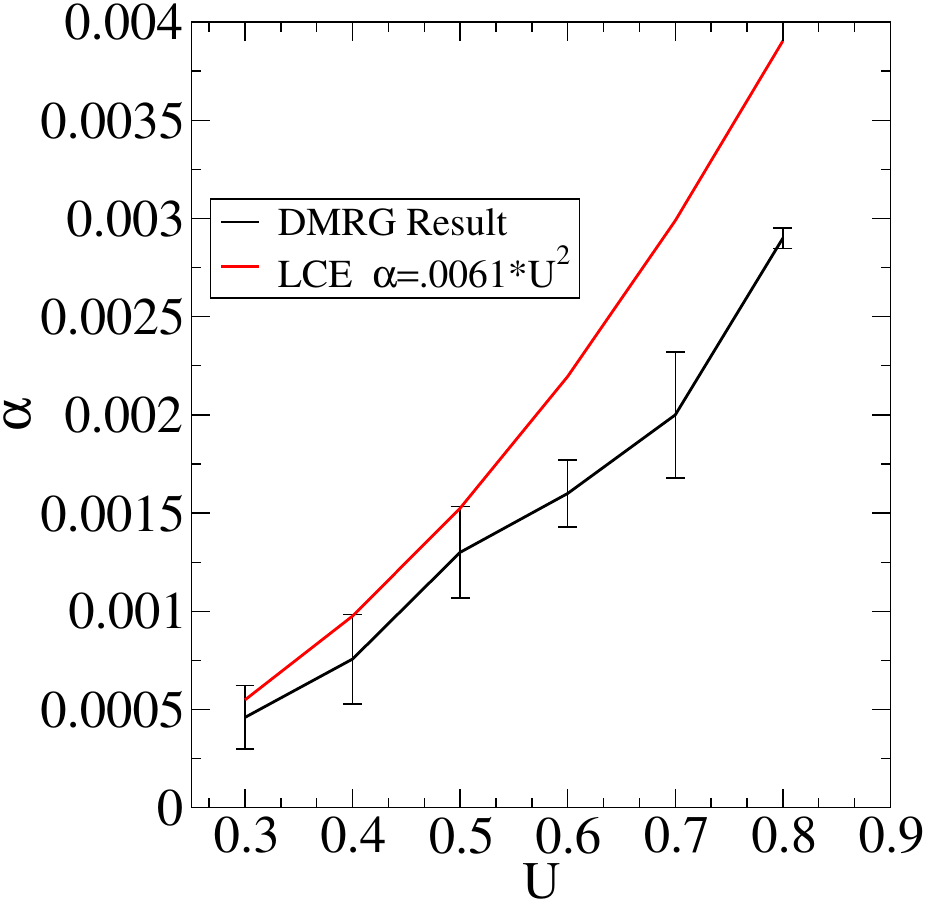}
\end{center}
\caption{\label{fig:exponent_sym_asym_green_function} (color online)  Inverse life-time of the impurity in antisymmetric sector (upper panel) and power-law exponent (lower panel) in the symmetric sector  at $t_\text{imp}= t_\text{b}=1$,  $\text{t}_\perp=1$,  $t_{\perp\text{imp}}=3$ as a function of $U$ at zero momentum. The black lines are numerical results and the red lines are LCE results, both numerical and analytical results show a nice agreement for small $U$.}
\end{figure}

For the parameters used in these two figures the gap in the antisymmetric sector is $\Delta_a=.33 t_b$. This value of the impurity-bath interaction corresponds to the regime of weak coupling for which a comparison with the analytical results of Sec.~\ref{sec:analytic} is meaningful. The comparison of the numerical results with the analytical results (\ref{eq:2eq30}) is shown in Fig.~\ref{fig:exponent_sym_asym_green_function}.
The numerical analysis thus fully confirms that in this regime the $G_s(0,t)$ and $G_a(0,t)$ decay as power-law and exponentially, respectively for large $t_{\perp\text{imp}}=3$. 
To further analyze the data we fit the numerical results to the form 
\begin{equation}
\begin{split}
 |G_{a}(p=0,t)| \propto \exp(- t/\tau(0))\\
  |G_{s}(p=0,t)| \propto \left(\frac{1}{t}\right)^{\alpha}
 \end{split}
\end{equation}
\begin{figure}
\begin{center}
 \includegraphics [ scale=0.4]{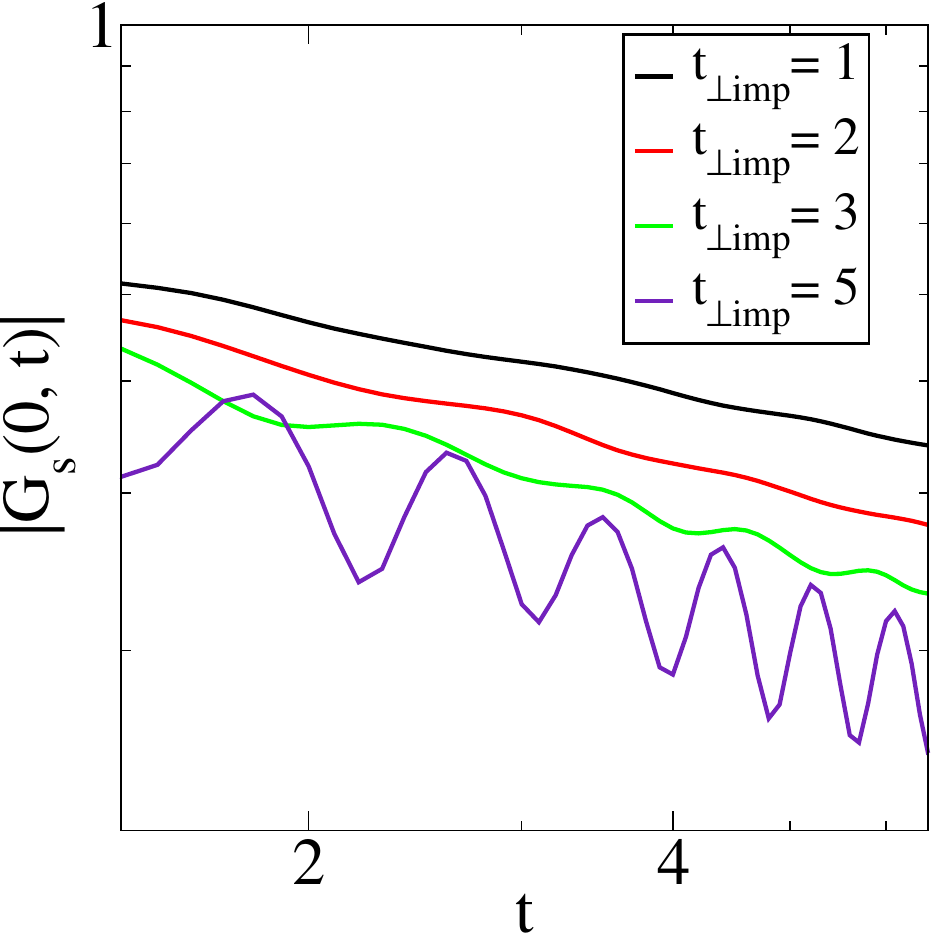}
 \end{center}
\caption{\label{fig:2fig4}
(color online)
Modulus of the Green's function of the impurity (see text) $|G_{s}(p,t)|$ at  momentum $p=0$. Parameters for the intra-chain hopping, inter-chain hopping (for the ladder), impurity hopping, impurity-bath interaction, impurity transverse tunneling in the ladder are $t_b=1$, $t_\perp=1$, $t_{\text{imp}} =1$, $U=\infty$, and $t_{\perp\text{imp}} =1,2, 3, 5$, respectively.}
\end{figure}

\subsubsection{Small U}
To analyze the data, we use the analytic estimates of Sec.~\ref{sec:analytic} which suggest a power law and an exponential decay of the Green's function of the symmetric and the antisymmetric mode respectively, at large $t_{\perp\text{imp}}$.

We fit the numerical data with the linked cluster expansion (LCE) result at $p=0$ and they agree very well with numerical results.  The numerical and analytical results for $G_{a}$ and $G_{s}$ are shown in Fig.~\ref{fig:exponent_sym_asym_green_function}.
\subsubsection{Hardcore bath-impurity repulsion}
\begin{figure}
\begin{center}
 \includegraphics[scale=0.35]{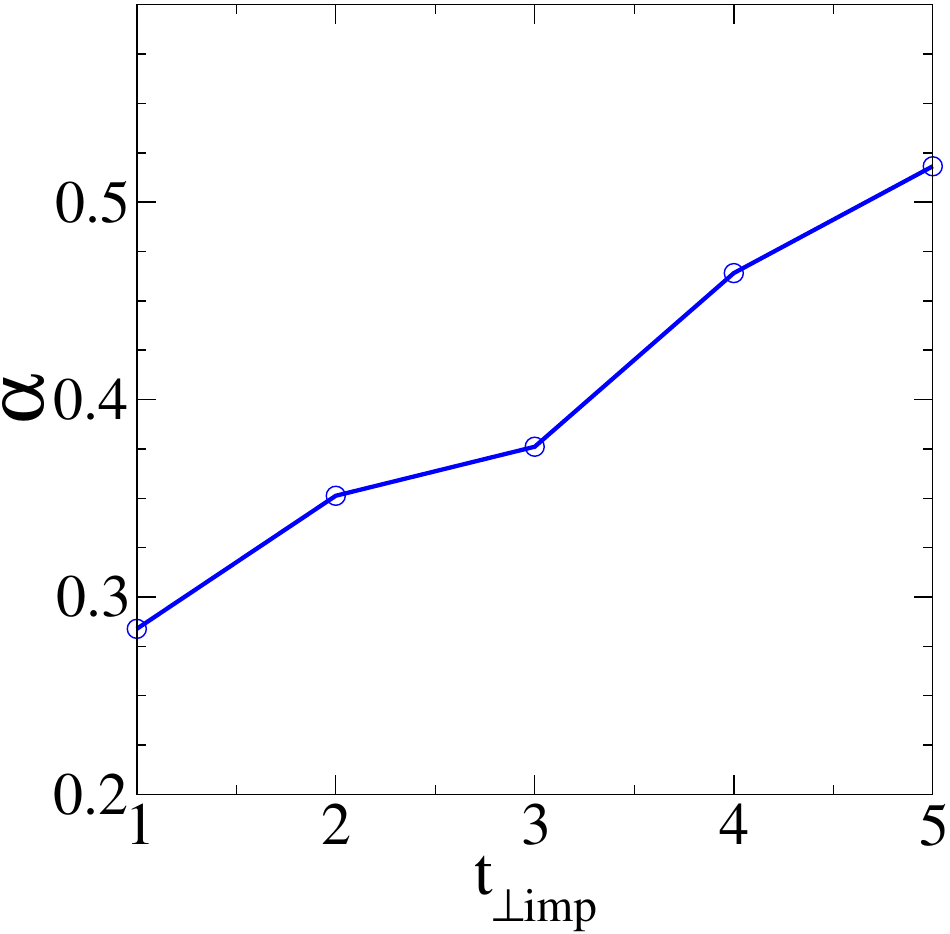}
\end{center}
\caption{\label{fig:2fig5} (color online) Power-law exponent of the symmetric Green's function of the impurity with a hardcore repulsion with the bath of hardcore bosons at filling $1/3$ as a function of $t_{\perp \text{imp}}$ at $t_{\text{imp}} = 1$, $U\rightarrow\infty$ and $\text{p=0}$. Circles are the numerical data for  $\chi=400$ and the line is a guide to the eyes.}
\end{figure}

In the case of $U \to \infty$, $|G_{s}(p, t)|$ is plotted on the log-log scale in Fig.~\ref{fig:2fig4} for different value of $t_{\perp\text{imp}}$ at $p=0$. As shown, $|G_{s}(p, t)|$ decays as a power-law and the power-law exponent as function of $t_{\perp\text{imp}}$ is depicted in Fig.~\ref{fig:2fig5}.
The power-law exponent increases as a function of $t_{\perp\text{imp}}$. 
For a small $t_{\perp\text{imp}}$, the exponent is similar to the one observed for the purely one-dimensional motion of an impurity in a two-leg ladder bath \cite{Naushad_Ladder_Impurity}, while for a large $t_{\perp\text{imp}}$, it is similar to the motion of an impurity in a one-dimensional bath ~\cite{Lamacraft_mobile_impurity_in_one_dimension,Adrian_Kantian_Mobile_Impuirty}.
\begin{figure}
\begin{center}
 \includegraphics [ scale=0.25]{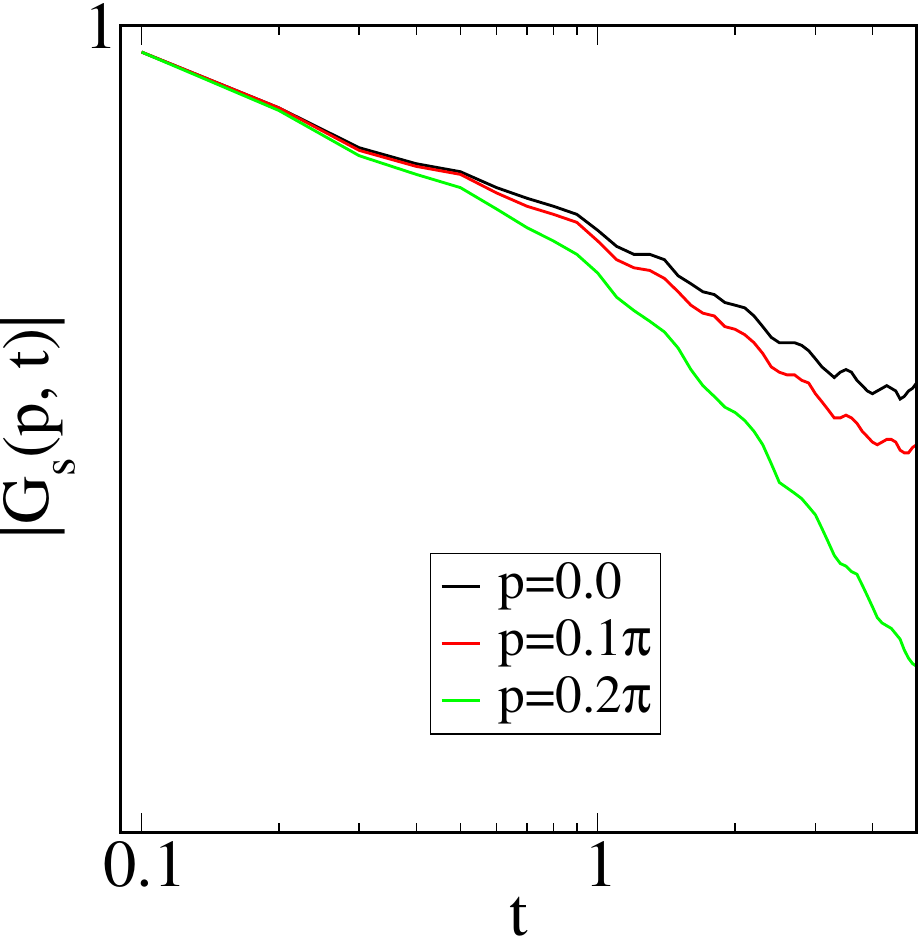}
  \includegraphics [ scale=0.25]
 {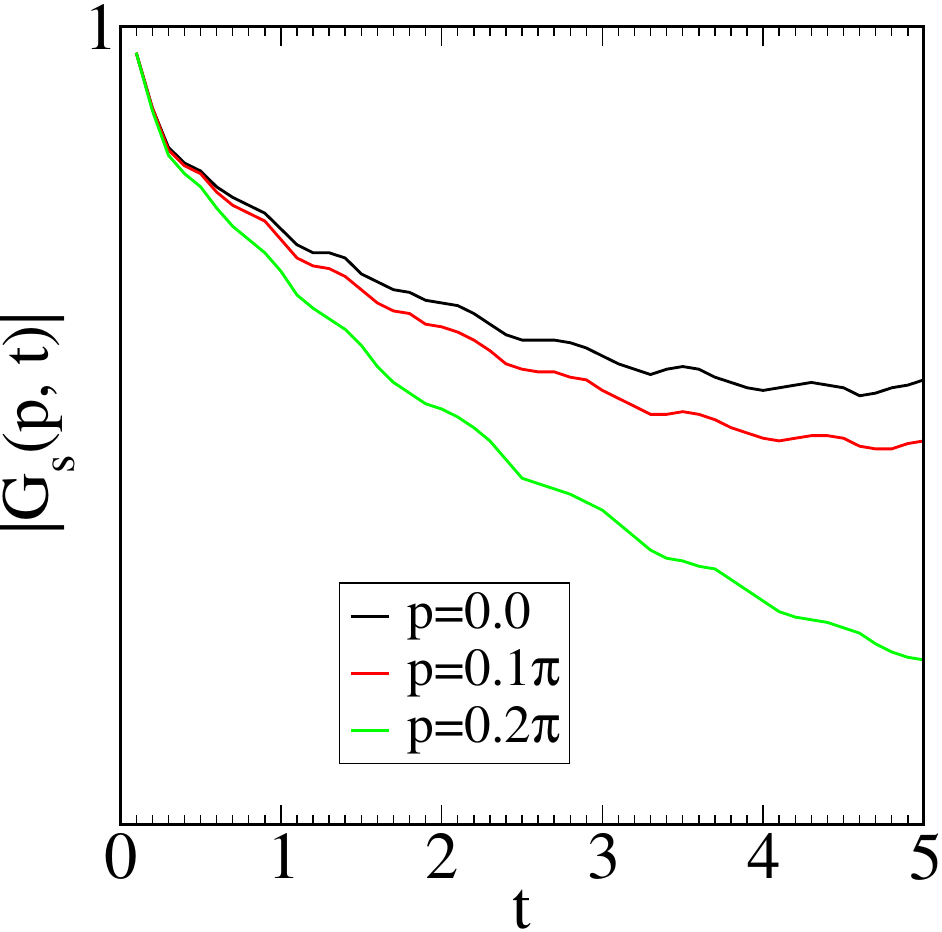}
  \includegraphics [ scale=0.25]{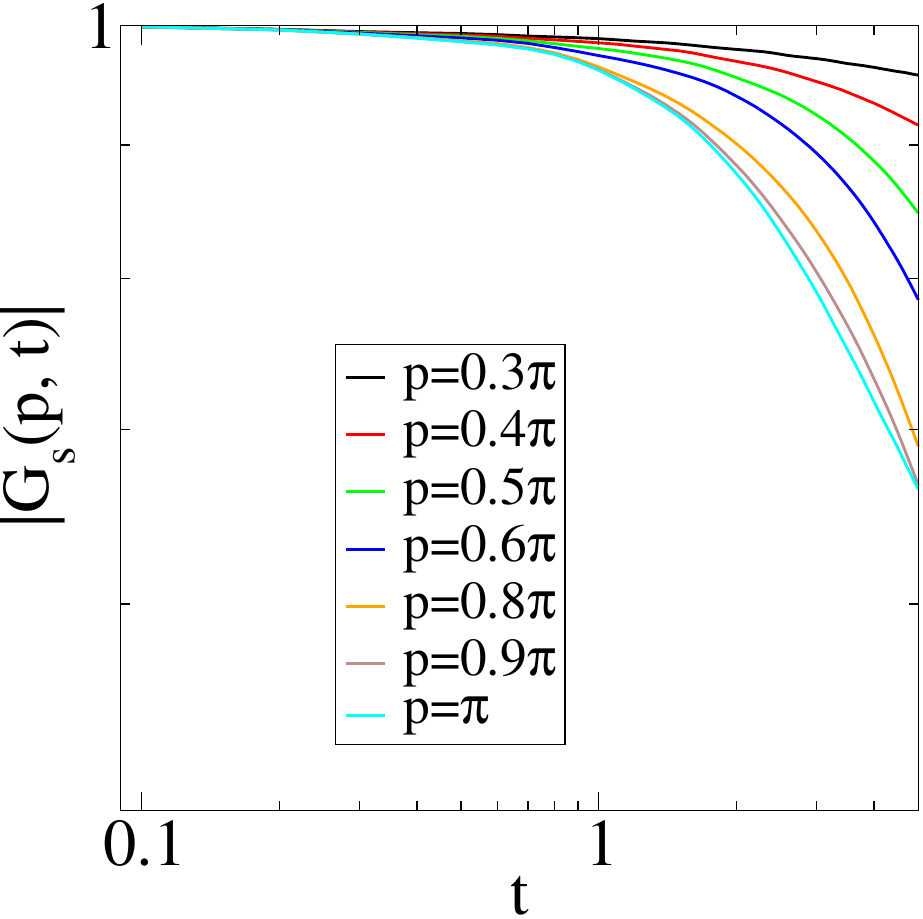}
  \includegraphics [ scale=0.25]
 {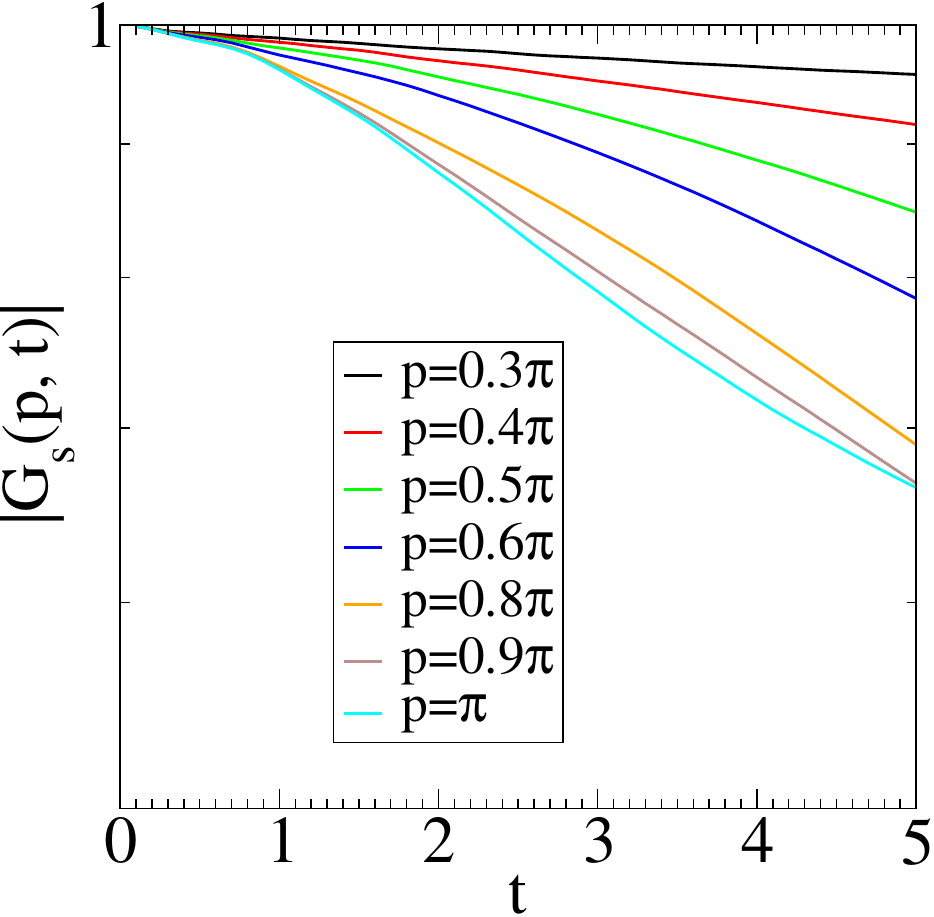}
 \end{center}
\caption{\label{fig:2fig2}
(color online)
Modulus of the Green's function of the impurity in the symmetric sector (see text) $|G_{s}(p,t)|$ at  different momenta $p$ (c.f. legend) for the impurity. The Hamiltonian parameters are $t_b=1$, $t_\perp=1$, $t_{\text{imp}} =1$,  $t_{\perp\text{imp}} =3$ and $U=1.0$ on log-log scale (left panel) and semi-log scale (right panel) . We observe a linear behavior on log-log scale for small momenta ($\text{p}=0-0.2\pi$), and linear behavior for large momenta $\text{p}=0.3\pi-\pi$ on semi-log scale. }
\end{figure}

\begin{figure}
\begin{center}
 \includegraphics [ scale=0.25]{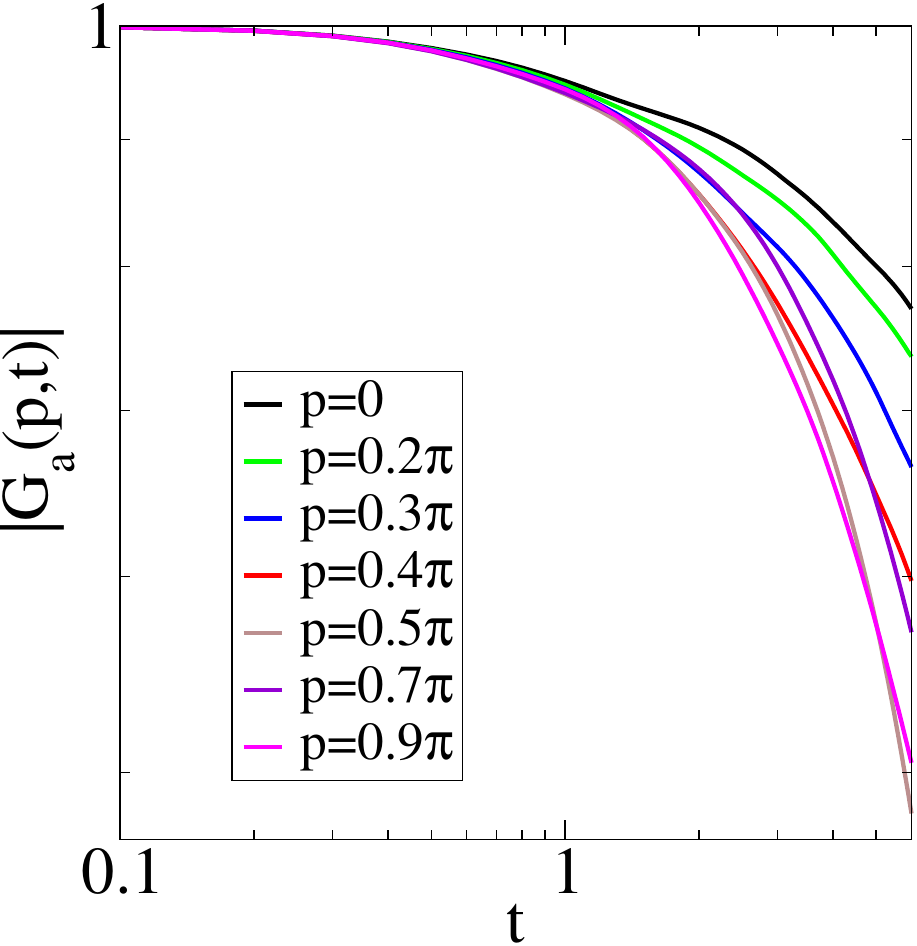}
  \includegraphics [ scale=0.25]
 {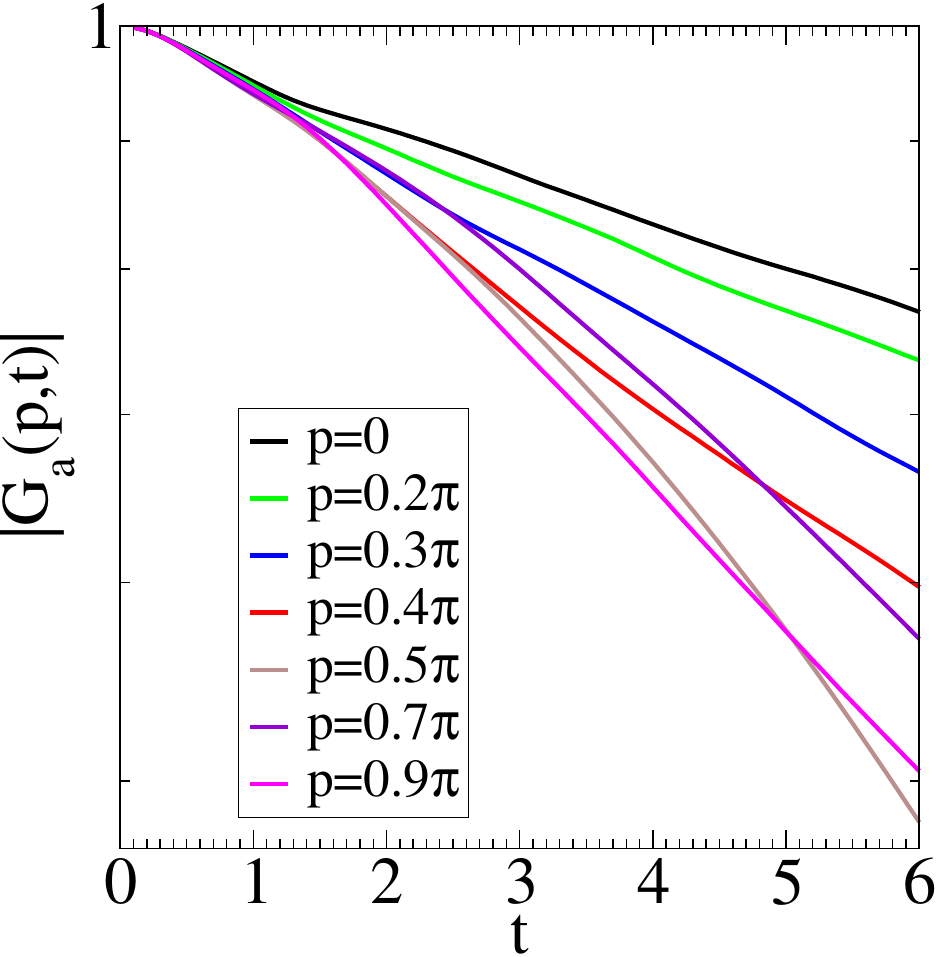}
 \end{center}
\caption{\label{fig:2fig3}Modulus of the Green's function of the impurity in antisymmetric sector (see text) $|G_{a}(p,t)|$ at  different momenta $p$ (shown in inset) for the impurity. The Hamiltonian parameters are $t_b=1$, $t_\perp=1$, $t_{\text{imp}} =1$,  $t_{\perp\text{imp}} =3$ and $U=1.0$ on log-log scale (left panel) and semi-log scale (right panel) . We observe a linear behavior on semi-log scale for all momenta ($\text{p}=0-0.9\pi$). }
\end{figure}

\begin{figure}
\begin{center}
 \includegraphics [ scale=0.4]{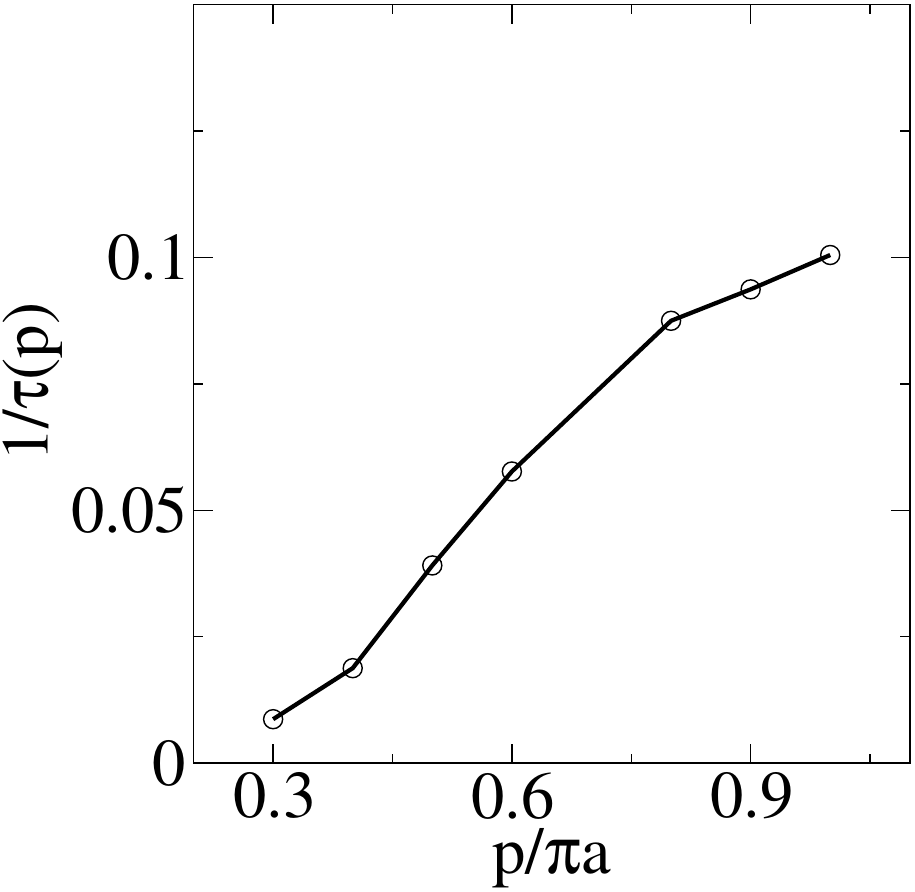}
  \includegraphics [ scale=0.4]
 {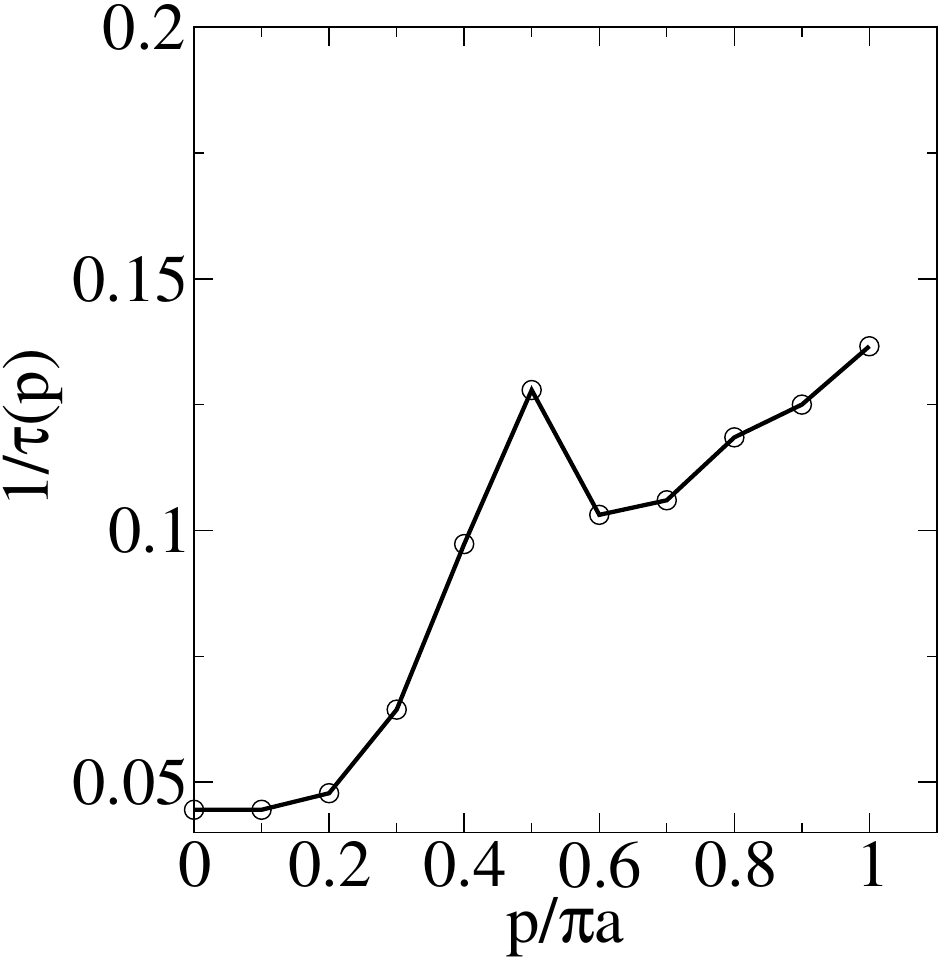}
 \end{center}
\caption{\label{fig:lifetime} Inverse life-time of the impurity in both symmetric (upper panel) and antisymmetric (lower panel) sectors as function of momentum at $t_b=1$, $t_\perp=1$, $t_{\text{imp}} =1$,  $t_{\perp\text{imp}} =3$ and $U=1.0$. }
\end{figure}
\blue{
\begin{figure}
\begin{center}
\includegraphics [ scale=0.3]
{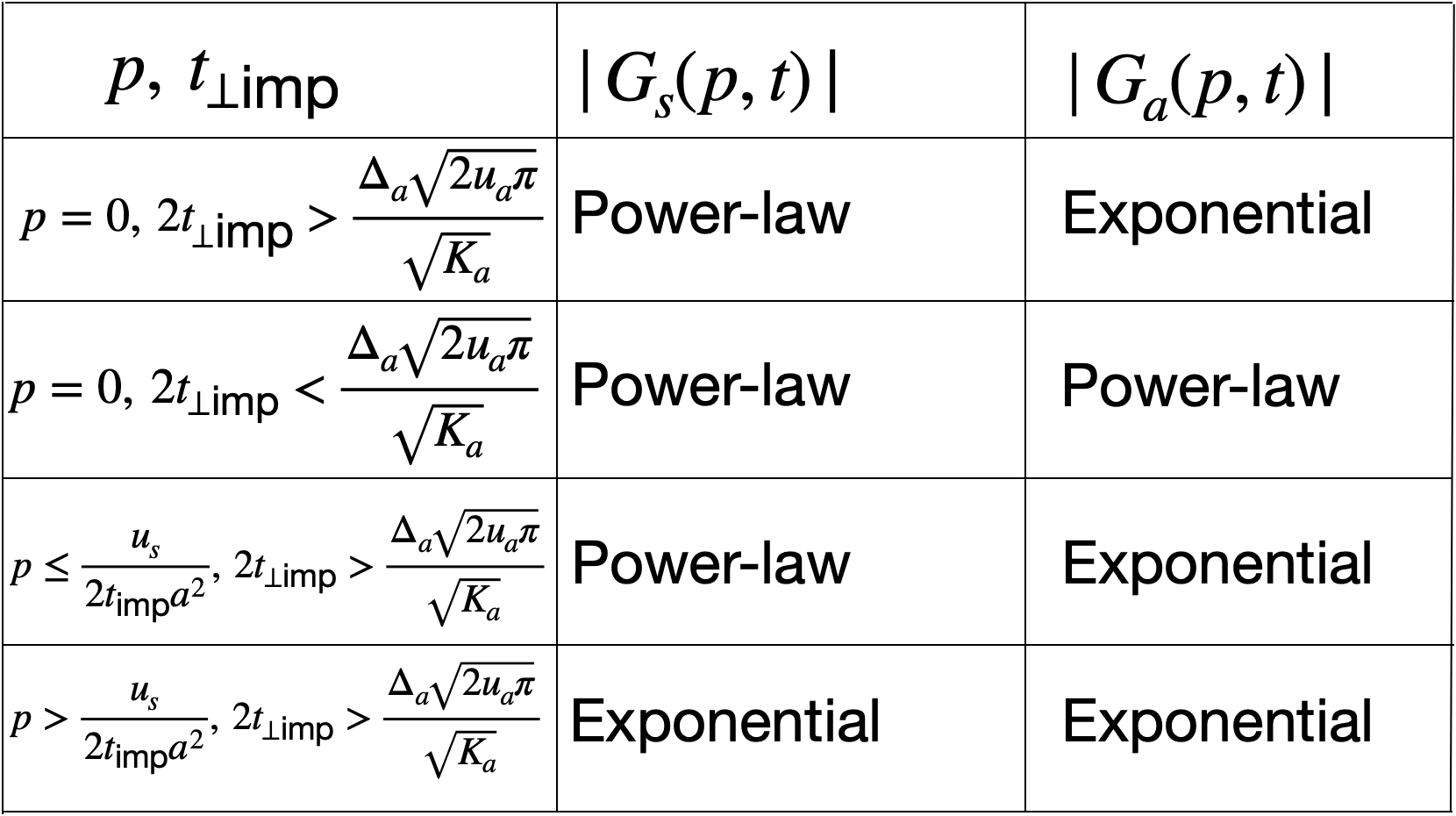}
 \end{center}
\caption{\label{fig:results_table} This table summarizes the nature of $G_s(p,t)$ and $G_a(p,t)$ for different momenta at small impurity-bath interaction.} 
\end{figure}
}
\subsection{Green's function at finite momentum} \label{sec:finitemoment}
As we have seen previously that the Green's function decays as a power-law in the symmetric mode and exponentially in the antisymmetric mode at $p=0$, now we turn on finite momentum. The Green's functions $|G_s(p,t)|$ and $|G_a(p,t)|$ for finite momentum are shown in Fig.~\ref{fig:2fig2} and \ref{fig:2fig3},  respectively,  for various momenta $p$ of the impurity. We find that $|G_s(p,t)|$ decays as power-law below $p=0.3\pi$. Beyond $p=0.3\pi$, it decays exponentially and the impurity in the symmetric mode enters into a  QP regime. An analogous  crossover has been established in the one-dimensional motion of an impurity in a one-dimensional bath ~\cite{Adrian_Kantian_Mobile_Impuirty} and two-leg bosonic ladder bath~\cite{Naushad_Ladder_Impurity,kamar2019quantum_thesis}.

The crossover depends on the TLL characteristics of the bath in the symmetric sector, namely the velocity of sound $u_s$ in the ladder and the TLL parameter $K_s$. Using the values extracted from \cite{crepin_bosonic_ladder_phase_diagram} we get
\begin{equation}
 p^* =\frac{u_s}{2 t_{\mathrm{imp}} a^2}=.93
\end{equation}
which is in reasonably good agreement with the observed change of behavior in Fig.~\ref{fig:2fig2}.

Beyond $\text{p}=\text{p}^*$ and for small $U$, the Green's function decays exponentially, the impurity behaves like a QP and the Green's function of the impurity in term of life time $\tau(p)$ is given by
\begin{equation}\label{eq:life_time}
\begin{split}
|G_{s}(p,t)|&=\exp(-t/\tau(p))
\end{split}
\end{equation}
In  the top panel of the Fig.~\ref{fig:lifetime}, we plot the inverse of life time $1/\tau(p)$ in the symmetric mode of the impurity, defined in eq. (\ref{eq:life_time})  of the QP as function of $\text{p}$ for interaction $U=1,  t_{\perp\text{imp}}=3$. As can be expected $1/\tau(p)$ increases with increasing $\text{p}$.

In Fig.~\ref{fig:2fig3}, we have shown $|G_a(p,t)|$ on semi-log and log-log scales, we find that the  $|G_a(p,t)|$  decays exponentially for all momenta at large $t_{\perp imp}=3.0$, and the impurity in the antisymmetric mode behaves like a QP. In the bottom panel of the Fig.~\ref{fig:lifetime}, we have shown the inverse lifetime as a function $p$, it shows a non-monotonic behavior but overall increases with increasing $p$.

In Fig.~\ref{fig:results_table}, we have summarized the results for small impurity-bath coupling.
\section{Discussion} \label{sec:discussion}
%Our findings suggest that the impurity exhibits very different dynamics in the ladder because of the motion in the horizontal and transverse directions than ones observed in one-dimensional (1D) motion of an impurity in a ladder and 1D bath.
Our findings suggest that the impurity exhibits very different dynamics in the ladder due to its motion in both the horizontal and transverse directions, compared to what is observed in the one-dimensional motion of an impurity in a ladder and a one-dimensional bath.

Let us first discuss the weak interaction limit. Initially, both the symmetric and antisymmetric modes of the impurity couple to both the gapless and gapped mode of the bath, but our numerical and analytical findings suggest that in the long time limit, the impurity in the antisymmetric sector effectively couples to  both gapped and gapless modes of the bath, and the impurity in the symmetric mode couples to the gapless mode of the bath with an effective interaction $U/\sqrt2$ (see \cref{eq:hambossym}). We have performed an LCE calculation at zero momentum by using the effective coupling between the impurity and the bath. In our LCE results, we find that Green's function of the impurity in the symmetric sector decays as a power-law. In contrast, in the antisymmetric sector, it decays exponentially for $2t_{\perp\text{imp}}>\frac{\Delta_a\sqrt{2 u_a \pi}}{\sqrt{K_a}}$ and as a power-law for $2t_{\perp\text{imp}}<\frac{\Delta_a\sqrt{2 u_a \pi}}{\sqrt{K_a}}$, the latter case is an emergent effect of the gap in the bath and transverse tunneling of the impurity on the dynamics of the impurity.
In order to explore the dynamics of impurities across a wide range of parameters, we have conducted a t-DMRG simulations.
For weak coupling, both numerical and analytical results show an excellent agreement as depicted in Fig.~\ref{fig:sym_asym_green_function} and  \ref{fig:exponent_sym_asym_green_function}.

The transverse tunneling of the impurity is the main ingredient in the exponential decay of the Green's function of the impurity in the antisymmetric mode while for the Green's function in the symmetric mode, the power-law exponent does not depend on the transverse tunneling of the impurity.
 
Now let us turn to the infinite interaction limit between the impurity and the bath.
We find that the Green's function in the symmetric sector decays as a power law at zero momentum, which is similar to the one observed in the impurity dynamics in a one-dimensional (1D) bath and two-leg ladder bath.

However, the power-law exponent increases with the tunneling amplitude of the impurity, which is in contrast to the behavior observed at small
U, where the exponent does not depend on $t_{\perp \text{imp}}$. We find that for small $t_{\perp \text{imp}}$, the power-law exponent is the same as that observed in 1D motion in the ladder bath \cite{Naushad_Ladder_Impurity} but for a larger value, the power-law exponent is equal to that observed in the 1D bath \cite{Adrian_Kantian_Mobile_Impuirty}. As a function of $t_{\perp \text{imp}}$, we observe that the impurity dynamics exhibits a dimensional crossover from ladder to 1D bath. For large transverse tunneling, one can understand that the impurity will energetically favor the symmetric mode of the impurity, and it would be hard to excite the impurity from symmetric to antisymmetric mode and vice versa. Hence the impurity effectively moves in a gapless 1D bath formed by the ladder's symmetric mode. This description contradicts our common understanding that in a ladder the power-law exponent should be smaller than that of 1D motion in the ladder~\cite{Naushad_Ladder_Impurity}.

It will be interesting to investigate how the dynamics of the impurity behaves with increasing the number of legs and this needs a further study. 

Our findings also suggest that the impurity in the symmetric sector at the zero momentum in the ladder can be viewed as an X-ray edge problem \cite{Giamarchi_bosonization}. The Green's function in the X-ray edge problem has similar behaviour  as the Green's function of the symmetric mode at zero momentum. 
Of course in this case, contrarily to the historical X-ray edge problem the impurity can move. 
At zero longitudinal tunneling of the impurity the impurity-ladder problem can be mapped into a spin-boson problem \cite{leggett_two_state} and by using an unitary transformation it can also be mapped into a Kondo problem \cite{spin-boson_model_to_Kondo_Luther}. The impurity-ladder problem can be viewed as a quantum simulator for the spin-boson model and Kondo problem. 

Now we finally turn to the case of finite momentum. For small interaction and small momentum, the Green's functions in the symmetric mode of the impurity decay as a power-law which are shown in the upper panel of the Fig.~\ref{fig:2fig2} . Beyond a critical momentum $\text{p}^\ast$, the Green's function is depicted in the lower panel of the Fig.~\ref{fig:2fig2}, it decays exponentially, and the impurity enters into a QP regime which is very similar to ones observed in the 1D motion of an impurity in a one-dimensional bath and in a two-leg bosonic ladder when $t_{\perp\text{imp}}$ is zero.  The critical momentum is precisely equal to that of 1D motion of an impurity in a ladder and in a 1D bath. 
The Green's functions of the impurity in the antisymmetric mode are depicted in Fig.~\ref{fig:2fig3}, and they always decay exponentially for all momenta, and the impurity in the antisymmetric mode always behaves like a QP. 

\section{Conclusion and perspectives} \label{sec:conclusion}

We have studied the dynamics of an impurity in a reservoir of hard-core bosons moving in a two-leg ladder where the  impurity may tunnel in both transverse and horizontal directions. We have computed the Green's function of the impurity for different momenta in order to understand the dynamics of the impurity. We use both analytical and numerical approaches, where later are comprised of the time-dependent DMRG.

The transverse tunneling of the impurity and gap in the bath drastically affects the impurity dynamics. Even in the antisymmetric sector, the impurity Green's function exhibits a power-law behavior for sufficiently small transverse tunneling of the impurity compared to the gap in the bath. For large impurity-bath coupling, the impurity dynamics exhibits a crossover from dynamics in a ladder bath to a 1D bath.

%When impurity-bath interactions are weak, the Green's function of the impurity in the symmetric sector decays as a power-law below a critical momentum and exponentially above the critical momentum like the 1D dynamics of an impurity in a two-leg ladder bath where transverse tunneling of the impurity is suppressed. However, in the antisymmetric sector the Green's function of the impurity always decays exponentially and the impurity behaves like a quasi-particle. 
%One can expect that when the bath is made of several 1D chains then in the lowest energy band of the impurity, the impurity would exhibit a crossover (beyond a critical momentum) from a power-law to an exponential decay. However, in the other energy sectors of the impurity, the Green's function would decay exponentially. 

%The above observations suggest  that the crossover of the dynamics of the impurity from a one-dimensional bath to a two-dimensional bath made up of finite number of 1D baths is not smoothly connected.  

The system we have studied can be tested experimentally in the context of circuit QED ~\cite{Diaz_Spin_Boson_Model, magazzu_spin_boson_model} and cold atoms. When the impurity  moves only in the transverse direction, the impurity acts like a two-level system which is analogous to a superconducting qubit, and the two-leg ladder bath acts like a one-dimensinal waveguide, and the impurity-reservoir interaction is the equivalent of the standard qubit-waveguide coupling. 
The bosonic ladder has been realized experimentally in ultracold gases ~\cite{atala_ladders_meissner,stoferle_tonks_optical} and atom chips~\cite{hofferberth_interferences_atomchip_LL} .The impurity dynamics in one-dimensional bath has been investigated experimentally using ultracold gases~\cite{Michael_Kohl_Spin_impurity_in_bose_gas,Minardi_Dynamics_of_impurities_in_one_dimension,Fukuhara_spin_impurity_in_one_dimension,meinert_bloch_oscillations_TLL}. Combination of these aspects and ongoing experimental advancement in the ultracold gases could provide the ideal testbed for our findings in near future.

\acknowledgments
Calculations were performed using the Matrix Product Toolkit~\cite{mptoolkit}.
We thank N. Laflorencie and G. Roux for providing us with the precise numerical value for the TLL parameters of the ladder of publication
\cite{crepin_bosonic_ladder_phase_diagram}. This work was supported in part by the Swiss National Science Foundation under grant 200020-219400 and ERC Starting Grant from the European Union’s Horizon 2020 research and innovation programme under grant agreement No. 758935; and the UK’s Engineering and Physical Sciences Research Council [EPSRC; grant number EP/W022982/1]. 
\appendix
\section{Green's function of the mobile impurity in the two-leg bosonic Ladder }\label{ap:LCE}
As we have shown in the main text (see section \ref{sec:analytic}) that for small interaction $U$ between the impurity and the ladder bath the impurity effectively coupled to the forward scattering terms of the gapped and gapless  modes of the bath.

In this section, we give a linked-cluster expansion (LCE) expression \cite{Mahan_Many_Particle_Physics} of the Green's function of the impurity.  We express  $\cos(\sqrt 2 \theta_a)=1-\theta_a^2$ and use the continuity equation 
\begin{align}
\nabla\phi_a(q,t)=\frac{\partial \theta_a(q,t)}{\partial t}.
\end{align} 
The impurity-bath Hamiltonian is expressed as
\begin{equation}
\begin{split}
H &=H_s+H_a+ H_\text{imp}+H_\text{coup},
\end{split}
\end{equation}
where $H_\text{s}$, $H_\text{a}$, $H_\text{imp}$, and $H_\text{coup}$ are the symmetric mode, antisymmetric mode of the bath, the impurity Hamiltonian, and the coupling between the impurity and the bath, respectively, and these terms in bosonized language are expressed as
\begin{equation}
\begin{split}
H_s &=\sum_{q}u_s|q| b^\dagger_{s, q}b_{s, q}, \\
H_a &= \frac{1}{2\pi}\int dx\Big[u_a K_a(\partial_x \theta_a)^2+\frac{u_a}{K_a} (\partial_x \phi_a)^2 \Big]\\
     & \blue{+}\Delta_a^2 \int dx \theta_a(x)^2, \\
%\sum_{q}\sqrt{v^2 q^2+\Delta_a^2} \theta_{a q}\theta_{a -q} \\
H_\text{coup} &=\sum_{{q}, {k}} \Big[V(q)(d^\dagger_{s,{ k}+{q}}d_{s,{ k}}+d^\dagger_{a,{ k}+{q}}d_{a,{ k}})(b_{{s, q}}+b^\dagger_{{s, -q}}) \\&
+\tilde U(d^\dagger_{a,{ k}+{q}}d_{s,{ k}}+d^\dagger_{s,{ k}+{q}}d_{a,{ k}})\frac{\partial \theta_a(q,t)}{\partial t}\Big], \\
H_\text{imp} &=\sum_{q}\epsilon_s(q) d^\dagger_{s,q}d_{s q}+\epsilon_a(q) d^\dagger_{a, q}d_{a, q},
\label{eq:3eq1B}
\end{split}
\end{equation}
where $\tilde U=-\frac{U}{\sqrt 2\pi}$, $b^\dagger$ and $d^\dagger$ are the creation operators for the bath in the bosonized language and the impurity respectively, $\epsilon_a(p)=-2t_{\text{imp}}\cos(p)+t_{\perp\text{imp}}$, $\epsilon_s(p)=-2t_{\text{imp}}\cos(p)-t_{\perp\text{imp}}$. $\Delta_a$ is the gap in the antisymmetric mode of the bath. The coupling term $V(q)$ can be expressed as
\begin{eqnarray}
V(q)&=&\frac{U}{\sqrt 2} \sqrt{\frac{K_s|q|}{2\pi L}}\exp\Big(-\frac{|q|}{2q_c}\Big).
\label{eq:2eq2A}
\end{eqnarray}
The Green's function of the impurity in symmetric and antisymmetric sectors are defined by
\begin{equation}
\begin{split}
G_{s}(p,t)&=-i\langle d_{s, p}(t)d^\dagger_{ s,p}(0)\rangle, \\
G_{a}(p,t)&=-i\langle d_{a, p}(t)d^\dagger_{ a, p}(0)\rangle .
\label{eq:2eq3B}
\end{split}
\end{equation}
By using LCE, (\ref{eq:2eq3B}) can be written as
\begin{eqnarray}
G_{s}(p,t)&=&-{i}e^{-i\epsilon_{s}(p)t}e^{F_{2s}(p,t)}\nonumber, \\
G_{a}(p,t)&=&-{i}e^{-i\epsilon_{a}(p)t}e^{F_{2a}(p,t)},
\label{eq:2eq4B}
\end{eqnarray}
where $F_{2, s/a}(p,t)$ is defined as
\begin{eqnarray}
F_{2 s/a}(p,t)&=&e^{i\epsilon_{s/a}(p)t}W_{2 s/a}(p,t).
\label{eq:2eq5B}
\end{eqnarray}
$W_{2 s/a}(p,t)$ is given by
\begin{eqnarray}
W_{2s/a}(p,t)&=&-\frac{1}{2}\int_{0}^{t} dt_1\int_{0}^{t} dt_2\nonumber\\&&\times \langle T_t d_{s/a,p}(t) H_{\text{coup}}(t_1)H_{\text{coup}}(t_2) d^\dagger_{s/a,p}(0)\rangle\nonumber.\\
\label{eq:2eq6B}
\end{eqnarray}
By employing the Wick's theorem $W_{2a}(p,t)$ can be expressed as
\begin{eqnarray}
\label{eq:2eq7B}
W_{2a}(p,t)&=-\sum_{q}\int_{0}^{t} dt_1\int_{0}^{t} dt_2 \Theta(t-t_1) \Theta(t_1-t_2) \nonumber \\&\times \Theta(t_2) \Big[V(q)^2 e^{-i\epsilon_a(p)(t-t_1)} e^{-i\epsilon_a(p+q)(t_1-t_2)} \nonumber\\& \times e^{-i\epsilon_a(p)t_2}e^{-i(u_s|q|(t_1-t_2))}\nonumber\\&+\frac{U^2}{4 \pi}\sqrt{K_a^2q^2+\frac{2\pi\Delta_a^2K_a}{u_a}}e^{-i\epsilon_a(p)(t-t_1)} \nonumber\\&\times e^{-i\epsilon_s(p+q)(t_1-t_2)} e^{-i\epsilon_a(p)t_2}\nonumber\\&\times e^{-i\Big(\sqrt{u_a^2q^2+\frac{2\pi\Delta_a^2u_a}{K_a}}(t_1-t_2)\Big)}\Big],
\end{eqnarray}
where $\Theta(t)$ is a step function, which is zero for $t<0$ and one for $t>0$. $\Theta(t)$ changes the limit of integration of $t_2$ and $t_1$, and $F_{2 a}(p,t)$ is modified as
\begin{align}\label{eq:2eq8B}
F_{2 a}(p,t)&=-\sum_{q}\int_{0}^{t} dt_1\int_{0}^{t_1} dt_2 \Big[ V(q)^2 e^{i\epsilon_a(p)t_1}
e^{-i\epsilon_a(p+q)(t_1-t_2)}\nonumber\\& \times e^{-i\epsilon_a(p)t_2}e^{-i(u_s|q|(t_1-t_2))}+ \frac{U^2}{4 \pi}\nonumber \\& \times \sqrt{K_a^2q^2+\frac{2\pi\Delta_a^2K_a}{u_a}} e^{i\epsilon_a(p)t_1}e^{-i\epsilon_s(p+q)(t_1-t_2)}\nonumber\\& \times e^{-i\epsilon_a(p)t_2}e^{-i\Big(\sqrt{u_a^2q^2+\frac{2\pi\Delta_a^2u_a}{K_a}}(t_1-t_2)\Big)}\Big].
\end{align}
We can simplify eq.~(\ref{eq:2eq8B}) as
\begin{equation}\label{eq:2eq9B}
\begin{split}
%F_{2 s}(p,t)&=-\sum_{q}\int du V(q)^2\int_{0}^{t} dt_1\int_{0}^{t_1} dt_2 \\& e^{it_1u} e^{-it_2 u}[\delta(u-\epsilon(p)+\epsilon(p+q)+v|q|)+\\& \delta(u-\epsilon(p)+\epsilon(p+q)+2t_{\perp\text{imp}}+v|q|)]\\
F_{2 a}(p,t)&=-\sum_{q}\int du \int_{0}^{t} dt_1\int_{0}^{t_1} dt_2 \\& \times \Big[V(q)^2 e^{it_1u} e^{-it_2 u} \delta(u-\epsilon(p)+\epsilon(p+q)+u_s|q|)\\&+\frac{U^2}{4 \pi}\sqrt{K_a^2q^2+\frac{2\pi\Delta_a^2K_a}{u_a}}\\&\times e^{-iu t_1} e^{iu t_2}\\& \times \delta\Big(u+\epsilon_a(p)-\epsilon_s(p+q)\\&-\sqrt{u_a^2q^2+\frac{2\pi\Delta_a^2u_a}{K_a}}\Big)\Big ].
\end{split}
\end{equation}
Finally, we integrate over $t_1$ and $t_2$ and the real part of $F_{2a}$ can be expressed as 
\begin{equation}
\begin{split}
\text{Re}[F_{2 a}(p,t)]&=-\sum_{q}\int du \Big[V(q)^2\frac{(1-\cos(u t))}{u^2} \\& \times \delta(u-\epsilon(p)+\epsilon(p+q)+u_s|q|)\\& + \frac{(1-\cos(tu))}{u^2} R_{2a}(u,p)].
\label{eq:2eq10B}
\end{split}
\end{equation}
Similarly one can show that 
\begin{equation}
\begin{split}
\text{Re}[F_{2 s}(p,t)]&=-\sum_{q}\int du   \\&\times\Big[V(q)^2\delta(u-\epsilon(p)+\epsilon(p+q)+u_s|q|)) \\&\times \frac{(1-\cos(u t))}{u^2}+ \frac{U^2}{4 \pi}\sqrt{K_a^2q^2+\frac{2\pi\Delta_a^2K_a}{u_a}}\\&\times\frac{(1-\cos(t u))}{u^2} \\&\times \delta\Big(u+\epsilon_s(p)-\epsilon_a(p+q)\\&-\sqrt{u_a^2q^2+\frac{2\pi\Delta_a^2u_a}{K_a}}\Big)
\Big],\\
&=-\int du  \Big[\frac{(1-\cos(u t))}{u^2} R_{1}(u,p)\\&+ \frac{(1-\cos(tu))}{u^2}R_{2s}(u,p)\Big].
\label{eq:2eq11B}
\end{split}
\end{equation}
$R_{1}(u)$, $R_{2a}$, and $R_{2s}$ are expressed as
\begin{equation}
\begin{split}
R_{1}(u, p) &= \frac{1}{2\pi}\int dq V(q)^2 \\& \times\delta(u-(\epsilon(p)-\epsilon(p+q)-u_{s}|q|)), \\
\end{split}
\label{eq:2eq15B}
\end{equation}
\begin{equation}
\begin{split}
R_{2a}(u, p) &= \frac{U^2}{8 \pi^2}\int dq \sqrt{K_a^2q^2+\frac{2\Delta_a^2K_a \pi}{u_a}} \\& \times \delta
\Big(u+2t_{\perp imp}+\epsilon(p)-\epsilon(p+q)\\&-\sqrt{u_a^2q^2+\frac{\Delta_a^2u_a 2\pi}{K_a}}\Big)\\
R_{2s}(u, p) &= \frac{U^2}{8 \pi^2}\int dq \sqrt{K_a^2q^2+\frac{2\Delta_a^2K_a \pi}{u_a}} \\& \times \delta
\Big(u-2t_{\perp imp}+\epsilon(p)-\epsilon(p+q)\\&-\sqrt{u_a^2q^2+\frac{\Delta_a^2u_a 2\pi}{K_a}}\Big).
\end{split}
\label{eq:2eq15Ba}
\end{equation}
For small $p$, $\epsilon(p)\simeq t_{\imp}p^2$ .
$R_1(u,p)$ is computed in Ref.~\cite{Adrian_Kantian_Mobile_Impuirty, Naushad_Ladder_Impurity} for $(p-\frac{u_s}{2t_{\text{imp}}})<0, \wedge  u<0$. However, the computation of $R_2(u,p)$ for an arbitrary $p$ is difficult analytically so we restrict at $p=0$. At $p=0$, $R_1$ is non-zero for $u<0$ can be expressed as
\begin{eqnarray} 
R_{1}(u, 0)&\propto&u,
\label{eq:2eq19B}
\end{eqnarray}
and for $u>\tilde{\Delta}-2t_{\perp imp}$
\begin{eqnarray} 
R_{2a}(u, 0)&\propto&A_2.
\label{eq:2eq19Ba}
\end{eqnarray}
Let us define $A_s(t)$ and $A_a(t) $ as
\begin{equation}
\begin{split}
A_a(t)&=-\frac{K_sU^2}{4 \pi^2 u_s^2}\int_{0} ^{\infty} du  \Big[\frac{1-\cos(u t)}{u}\Big]\\&+\int_{\frac{\Delta_a\sqrt{2 u_a \pi}}{\sqrt{K_a}}-2t_{\perp\text{imp}}} ^{\infty} du\Big[\frac{1-\cos(u t)}{u^2}  A_2\Big], \\
A_s(t)&=-\frac{K_sU^2}{4 \pi^2 u_s^2}\int_{0} ^{\infty} du  \Big[\frac{1-\cos(u t)}{u}\Big]\\& +\int_{\frac{\Delta_a\sqrt{2 u_a \pi}}{\sqrt{K_a}}+2t_{\perp\text{imp}}} ^{\infty} du\Big[\frac{1-\cos(u t)}{u^2}  R_{2s}(u,0)\Big].
\end{split}
\label{eq:eq_new1}
\end{equation}
In the long time limit, $A_s$ and $A_a$ can be expressed as 

\begin{equation}
\begin{split}
A_a(t)&\simeq -A_2 \pi t\\&-
\frac{K_sU^2}{4 \pi^2 u_s^2}\log(t), \mathrm{if} \ 2t_{\perp\text{imp}}>\frac{\Delta_a\sqrt{2 u_a \pi}}{\sqrt{K_a}} \\
&\simeq -
\frac{K_sU^2}{4 \pi^2 u_s^2}\log(t), \mathrm{if} \ 2t_{\perp\text{imp}}<\frac{\Delta_a\sqrt{2 u_a \pi}}{\sqrt{K_a}} \\
%&\simeq -\frac{A_2 \pi t}{2} \\&-
%\frac{K_sU^2}{4 \pi^2 u_s^2}\log(t), \mathrm{if} \ 2t_{\perp\text{imp}}=\frac{\Delta_a\sqrt{2 u_a \pi}}{\sqrt{K_a}}\\
%&\simeq -A_2 \pi t \Theta\Big(2t_{\perp\text{imp}}-\frac{\Delta_a\sqrt{2 u_a \pi}}{\sqrt{K_a}}\Big)\\&-
%\frac{K_sU^2}{4 \pi^2 u_s^2}\Big(1+\frac{12 t_{\text{imp}}^2p^2}{u_s^2}\Big)\log(t) \\
%-\text{Ci}[u t] +\log[u]+A_2/(-2 t_{\perp\text{imp}} + 
  %u) \\& (-2 t_{\perp\text{imp}} + 
  % 2 t_{\perp\text{imp}} \cos[t (2 t_{\perp\text{imp}} - u)] - \\& (-2 t_{\perp\text{imp}} + u) \text{Ci}[t (-2 t_{\perp\text{imp}} + u)] - 
 %  2 t_{\perp\text{imp}} \\& \log[-2 t_{\perp\text{imp}} + u] + u \log[-2 t_{\perp\text{imp}} + u] + \\&
  % 2 t_{\perp\text{imp}} t (2 t_{\perp\text{imp}} - u) \text{Si}([2 t_{\perp\text{imp}} t - u t])\\%+A_1(t \text{Si}(u t -2 t_{\perp\text{imp}}t)+\frac{\cos(u t -2 t_{\perp\text{imp}}t)-1}{u  -2 t_{\perp\text{imp}}})\\
A_s( t)&\simeq -\frac{K_sU^2}{4 \pi^2 u_s^2}\log(t),
%-\text{Ci}[u t] +\log[u]+A_2/(2 t_{\perp\text{imp}} + 
%  u) \\& (2 t_{\perp\text{imp}} 
 %  -2 t_{\perp\text{imp}} \cos[t (2 t_{\perp\text{imp}} + u)] - \\& (2 t_{\perp\text{imp}} + u) \text{Ci}[t (2 t_{\perp\text{imp}} + u)]+ 
%   2 t_{\perp\text{imp}} \\& \log[2 t_{\perp\text{imp}} + u] + u \log[2 t_{\perp\text{imp}} + u] + \\&
 %  2 t_{\perp\text{imp}} t (2 t_{\perp\text{imp}} + u) \text{Si}([2 t_{\perp\text{imp}} t + u t])\\&%+A_1(t \text{Si}(u t +2 t_{\perp\text{imp}}t)+\frac{\cos(u t +2 t_{\perp\text{imp}}t)-1}{u  +2 t_{\perp\text{imp}}})
\end{split}
\label{eq:eq_new2}
\end{equation}

%Where $A_2=\frac{U^2 u_a}{8 \pi^2 K_a}\frac{(-u_a/(2 t_{\text{imp}})+f_1)^2}{\sqrt{(-u_a/(2 t_{\text{imp}})+f_1)^2+\tilde{\Delta}_a^2/u_a^2} f_1}$, $f_1=\sqrt{\frac{u_a^2}{4  t_{\text{imp}}^2}+\frac{\tilde{\Delta}_a+2 t_{\perp\text{imp}}}{t_{\text{imp}}+\tilde{\Delta}_a^2/u_a^2}}$, $\tilde{\Delta}_a=\frac{\Delta_a\sqrt{u_a2\pi}}{\sqrt{K_a}}$.
where $\Theta(x)$ is a Heaviside step function defined as $\Theta(x)=1$ for $x> 0$ and $0$ for $x<0$ and $A_2$ is expressed as
\begin{equation}
A_2\simeq\frac{U^2 K_a}{4 u_a \pi^2}\frac{(u_a^2q_-^2+\tilde{\Delta}^2)}{ q_-(2t_{\text{imp}}\sqrt{u_a^2q_-^2+\tilde{\Delta}^2}+u_a^2)},
\end{equation}

and $\tilde{\Delta}$ and $q_-$ are expressed as
\begin{equation}
\begin{split}
\tilde{\Delta}&=\frac{\Delta_a\sqrt{u_a2\pi}}{\sqrt{K_a}},\\
q_-&=\sqrt{\frac{2 t_{\perp\text{imp}}}{t_{\text{imp}}}+\frac{u_a^2}{2 {t_{\text{imp}}^2}}-\sqrt{\Big(\frac{2 t_{\perp\text{imp}}}{t_{\text{imp}}}+\frac{u_a^2}{2 {t_{\text{imp}}}^2}\Big)^2-\frac{(4 t_{\perp\text{imp}}^2-\tilde{\Delta}^2)}{t_{\text{imp}}^2}}}.
\end{split}
\end{equation}
%where $A_2=\frac{2(u_a^2q_-^2+\tilde{\Delta}^2)}{2 t_{\text{imp}}q_-\sqrt{u_a^2q_-^2+\tilde{\Delta}^2}+u_a^2}$, $\tilde{\Delta}=\frac{\Delta_a\sqrt{u_a2\pi}}{\sqrt{K_a}}$, and $q_-=\sqrt{\frac{2 t_{\perp\text{imp}}}{t_{\text{imp}}}+\frac{u_a^2}{2 {t_{\text{imp}}^2}}-\sqrt{(\frac{2 t_{\perp\text{imp}}}{t_{\text{imp}}}+\frac{u_a^2}{2 {t_{\text{imp}}}^2})^2-(4 t_{\perp\text{imp}}^2-\tilde{\Delta}^2)}}$.
By using equations (\ref{eq:2eq10B}, \ref{eq:2eq11B}, \ref{eq:eq_new1}, \ref{eq:eq_new2}), the final expression of $F_{2 s}, F_{2 a}$ in long time limit is given by 
\begin{eqnarray}
\text{Re}[F_{2 a}(0,t)]&\simeq A_a(t)
\label{eq:2eq20B}
\end{eqnarray}
\begin{eqnarray}
\text{Re}[F_{2 s}(0,t)]&\simeq A_s(t)
\label{eq:2eq21B}
\end{eqnarray}

leading to the Green's functions decay as
\begin{equation}
|G_{a}(0,t)|= e^{A_a(t) },
\label{eq:2eq22B}
\end{equation}
\begin{equation}
|G_{s}(p,t)|=e^{A_s(t)}.
\label{eq:2eq23B}
\end{equation}
\bibliography{totphys-A-J,totphys-K-Z,Impurity_in_Ladder,impurity_TG}
\end{document}